\documentclass[11pt]{article}

\usepackage{arxiv}
\usepackage{graphicx}
\usepackage{subcaption}
\usepackage{amsmath,amssymb}
\usepackage{booktabs}
\usepackage{multirow}
\usepackage{float}
\usepackage{hyperref}
\usepackage{url}
\usepackage[numbers]{natbib}

\graphicspath{{./}}

\renewcommand{\headeright}{}
\renewcommand{\undertitle}{}
\renewcommand{\shorttitle}{}

\title{Adapting Automotive Aerodynamics Surrogates to New Vehicle Families via Transfer Learning}
\author{
  Seunghwan Keum \quad Alok Warey \\
  General Motors Research and Development
}
\date{May 2026}

\begin{document}
\maketitle

\begin{abstract}
Deploying Scientific Machine Learning surrogates in industrial Computational Fluid Dynamics (CFD) workflows requires adapting pretrained models to new vehicle families without large datasets; yet whether geometric representations learned by a geometry encoder transfer to topologically distinct shapes remains unvalidated.

We address this through leave-one-family-out experiments on a 61.47\,million-parameter Transformer surrogate (AB-UPT) pretrained on four vehicle families (411 external aerodynamics cases) and adapted to the held-out fifth with only 20 samples.
Three strategies are compared: Full Fine-Tuning (FFT), Lightweight Fine-Tuning (LFT), and Low-Rank Adaptation (LoRA).
The central finding is that pretrained geometry encoders learn transferable representations, but the adaptation mechanism determines whether they can be exploited.
FFT destabilizes as 61.47\,million unconstrained parameters overfit to 20 samples ($R^2\!=\!0.40$); LFT fails because the frozen encoder cannot represent unseen shapes ($R^2\!<\!0$).
LoRA resolves both: rank-constrained adapters injected into all layers regularize the loss landscape while preserving pretrained features, achieving $R^2\!=\!0.85 \pm 0.02$ across all five families with 50\,\% lower force RMSE than FFT and 28\,\% lower pointwise field errors.
LoRA also outperforms from-scratch training using 3$\times$ more target-family data, eliminating the need for large per-family datasets.
These results recast LoRA from a memory-saving convenience into a \emph{convergence enabler} for geometry transfer: a shared backbone paired with lightweight per-family adapters trainable in hours from minimal data.
\end{abstract}

\section*{Research Highlights}
\begin{itemize}
  \item This study provides the first empirical validation of geometry encoder transferability in an industrial-scale Transformer surrogate (61.47\,million parameters), systematically comparing Full Fine-Tuning, Lightweight Fine-Tuning, and Low-Rank Adaptation for adapting to unseen vehicle families.
  \item Conventional weight-modification strategies, Full Fine-Tuning and Lightweight Fine-Tuning, exhibit poor transferability: FFT suffers from training instability due to an ill-conditioned loss landscape with 3\,million+ parameters per sample, while LFT fails entirely because the frozen geometry encoder cannot represent unseen shapes.
  \item Low-Rank Adaptation achieves robust geometry transfer ($R^2 = 0.85 \pm 0.02$) with only 20 training samples. Beyond its well-known role as a memory-efficient fine-tuning method, LoRA serves as a \emph{training convergence enabler}: the rank constraint restricts gradient updates to a low-dimensional subspace, regularizing the loss landscape and preventing the catastrophic overfitting that destabilizes unconstrained fine-tuning in the extreme low-data regime.
  \item Domain consistency between pretraining and fine-tuning, particularly normalization statistics and coordinate-embedding alignment, is identified as a critical but previously overlooked prerequisite for geometry transfer; violating this requirement degrades predictions to worse-than-random levels.
\end{itemize}

\section{Introduction}\label{sec:introduction}

Artificial intelligence is reshaping many aspects of modern engineering.
At the center of recent breakthroughs, from conversational agents to multimodal reasoning, lies the Transformer architecture~\cite{vaswani2017attention}, which has expanded from Natural Language Processing (NLP) into computer vision, scientific computing, and numerous other domains.
A natural frontier is Scientific Machine Learning (SciML), where deep learning is used to approximate the solutions of Partial Differential Equations (PDEs) governing physical phenomena such as fluid dynamics, structural mechanics, and heat transfer.

Significant efforts have been made to develop and apply SciML methods specifically for external aerodynamics, spanning both aviation and ground vehicles.
Several large open-source databases have been released, including the AirfRANS dataset~\cite{airfoil_dataset} and the DrivAer family of automotive benchmarks~\cite{drivaerml,driveaermlpp,drivaerstar}.
On the architectural front, Convolutional Neural Networks (CNNs) were among the first approaches applied to vehicle aerodynamics~\cite{sae2026010600} but showed limited success due to their inherent requirement for constant-distance pixel or voxel grids, an unnatural representation for complex 3D geometries.
Graph Neural Networks (GNNs)~\cite{gilmer2017neural,pfaff2021learning} improved on this by representing the computational mesh directly as a graph, yet their application to industrial Computer-Aided Engineering (CAE) has remained challenging due to the sheer volume of mesh points, often tens to hundreds of millions, that arises in production-level simulations.
Furthermore, GNNs rely on message passing over multiple hops for long-range interactions, making it difficult to capture phenomena such as the influence of front grille design on the vehicle wake~\cite{ashton2025fluid}.
Neural Operators were subsequently proposed as an alternative paradigm, learning functional mappings via Green's function kernels~\cite{subramanian2023foundation}; extensions such as Fourier Neural Operator (FNO)~\cite{li2021fno}, Geometry-Informed Neural Operator (GINO)~\cite{li2023gino}, and DoMINO~\cite{ranade2025domino} followed, though the long-range interaction challenge persists.
The Transformer architecture~\cite{vaswani2017attention} has emerged as the most attractive alternative, as its self-attention mechanism evaluates pairwise relationships among all elements simultaneously, eliminating the multi-hop limitation of GNNs.
Applications to SciML began with the Universal Physics Transformer (UPT)~\cite{alkin2025universalphysicstransformersframework}, later expanded into AB-UPT~\cite{alkin2025abupt} to address memory scalability, and independently with the Transolver~\cite{transolver2025} and its successors~\cite{transolver2025pp,Transolver3,geotransolver}.
Comprehensive benchmarks confirm that Transformer-based surrogates outperform other architectures in prediction accuracy~\cite{elrefaie2025carbenchcomprehensivebenchmarkneural}, while memory and computational cost at inference remain open challenges.

With SciML architectures maturing rapidly, the practical bottleneck shifts from model design to \emph{training data generation}.
Unlike NLP and computer vision, where data can be scraped or crowdsourced at scale, SciML training data must be actively generated through computational simulations, each requiring a carefully designed CAE campaign that demands domain expertise, high-performance computing resources, and commercial solver licenses~\cite{ashton2025fluid}.
Consequently, large-scale data generation has been limited and the open-source databases have been indispensable resources for the research community~\cite{drivaerml}.
However, the vehicle types encountered in industry are far more diverse than what open-source databases cover, spanning from small crossovers to large SUVs and pickup trucks, and design cycles are growing shorter with significant shape changes within the same segment over just a few years.
Covering multiple vehicle families across successive design generations would require training data volumes that may well exceed what is typically needed for the actual development CAE runs, yielding a poor return on investment for surrogate development.
For the broad industrial adoption of SciML, a training methodology that requires minimal data for new vehicle families is therefore critical.

Transfer learning, leveraging knowledge learned from one set of geometries to adapt to new, unseen shapes with minimal additional data~\cite{zhuang2020comprehensive}, is a promising approach to this data efficiency challenge.
Recent reviews have identified transfer learning as a key enabler for scaling machine learning methods in engineering simulation~\cite{pan2025piml_jcise,nguyen2023ml_multiscale}, yet the question of whether pretrained geometric representations can generalize across geometrically distinct shapes remains largely unexplored.
A SciML surrogate that cannot adapt to new shapes without costly retraining is of limited practical value, yet a fundamental question remains unanswered: \emph{Can the geometric representations learned by an explicit geometry encoder transfer to geometrically distinct, unseen shapes?}
This study addresses this question and, conditional on an affirmative answer, asks a second: \emph{Which fine-tuning strategy is most effective for geometry adaptation in the low-data regime?}

To this end, we apply and compare multiple transfer learning strategies, namely FFT, LFT, and LoRA, to an industrial CAE dataset of ground vehicle external aerodynamics, deliberately limited in sample size to reflect realistic industrial conditions.
The surrogate model learns to predict pointwise surface pressure and wall shear stress (friction force) distributions from vehicle geometry alone.
These surface field predictions serve a dual purpose: they are integrated over the vehicle body to yield the scalar aerodynamic drag force, the primary engineering metric in production vehicle development, and, more importantly, they reveal the \textit{spatial distribution} of drag contributions across the vehicle surface.
It is this spatial information that enables design engineers to identify which geometric features (e.g., A-pillar angle, underbody geometry, rear spoiler shape) contribute most to drag, guiding targeted design improvements.
Alternative architectures such as PointNet~\cite{qi2017pointnet}, Graph Attention Networks~\cite{velickovic2018gat}, or Global Graph Transformers~\cite{kong2023goat} could in principle predict a single scalar drag coefficient directly from the input point cloud; however, such approaches provide no information about \emph{where} on the surface the drag originates, limiting their utility for iterative design refinement.
The ability to predict full surface field distributions is therefore a non-negotiable requirement for the intended design application.
This study accordingly adopts an advanced Transformer architecture capable of predicting complete surface pressure and friction fields, enabling rapid design iteration through spatially resolved aerodynamic feedback.
To address the industrial reality that development programs span multiple vehicle families simultaneously, we systematically evaluate the transfer learning capability of this architecture: whether a model pretrained on existing families can be efficiently adapted to a new, topologically distinct family with minimal additional data.

The remainder of this paper is organized as follows.
Section~\ref{sec:related_work} reviews transfer learning methods and their application in SciML, identifying the specific research gap this study addresses.
Section~\ref{sec:methodology} describes the baseline Transformer architecture, the vehicle aerodynamics dataset, and the experimental design.
Section~\ref{sec:results} presents the quantitative results, and Section~\ref{sec:discussion} discusses their implications, limitations, and directions for future work.

\section{Related Work}\label{sec:related_work}

\subsection{Transfer Learning in Scientific Machine Learning}

Transfer learning adapts a model trained on a source domain to a target domain with limited data, and has become standard practice in computer vision and NLP~\cite{zhuang2020comprehensive,yosinski2014transferable}.
Wang et al.~\cite{wang2025tlpinn} conducted the most directly relevant prior study, systematically comparing full fine-tuning, lightweight fine-tuning (analogous to our LFT), and LoRA for Physics-Informed Neural Networks (PINNs) across different boundary conditions, materials, and geometries.
Their key findings were that LoRA and full fine-tuning significantly improve convergence speed and accuracy, while lightweight fine-tuning performs poorly in most scenarios, because PINNs, unlike CNNs, do not exhibit hierarchical features that would justify freezing early layers.
For geometry transfer specifically (circular to elliptical plate holes), LoRA achieved the highest accuracy, and the optimal rank was found to scale with the dissimilarity between source and target domains.
However, their experiments were limited to 2D settings with shallow four-layer MLPs ($\sim$30,000 parameters) that lack any explicit geometry encoder; whether these findings generalize to architectures with dedicated geometry encoders, where the encoder's learned representations must themselves transfer, and to industrial-scale Transformer surrogates ($\sim$60\,million parameters) operating on complex 3D geometries, remains unvalidated.
The present study addresses both questions.

Full Fine-Tuning (FFT) updates all parameters of the pretrained model on the new target data.
This is the most common approach in SciML transfer learning studies~\cite{subramanian2023foundation,leng2024varfidelity}, but key risks include catastrophic forgetting~\cite{kirkpatrick2017ewc} and optimization instability in the low-data regime, where the vast number of trainable parameters creates an ill-conditioned loss landscape.
Lightweight Fine-Tuning (LFT) freezes the backbone and trains only the last few layers or a task-specific head~\cite{yosinski2014transferable}, assuming that early layers capture general features.
However, this assumption may not hold when the target is a geometrically distinct shape family that differs fundamentally from the training distribution, since the frozen layers may not have learned sufficiently general features for the new geometry.

Parameter-Efficient Fine-Tuning (PEFT) methods~\cite{ding2023peft_survey,lialin2023scaling} inject a small number of trainable parameters into the frozen pretrained model.
Adapter modules insert small trainable Multi-Layer Perceptron (MLP) blocks between frozen layers~\cite{zhang2023llama_adapter}; more recently, Zhang et al.~\cite{zhang2026fadapter} proposed F-Adapter, a frequency-adaptive variant designed specifically for scientific machine learning that outperforms standard LoRA on neural operator benchmarks by respecting the spectral structure of PDE solutions.
LoRA~\cite{hu2021lora} is the most widely adopted PEFT approach, decomposing weight updates into low-rank matrices at each layer.
LoRA has become the dominant fine-tuning paradigm for billion-scale language models, with extensions including adaptive rank allocation~\cite{zhang2023adalora}, quantization-aware training~\cite{dettmers2023qlora}, and weight decomposition~\cite{liu2024dora}.
Table~\ref{tab:finetuning_overview} provides a summary overview of these strategies.

\begin{table}[ht]
  \centering
  \caption{Overview of fine-tuning strategies for pretrained models.}
  \label{tab:finetuning_overview}
  \small
  \begin{tabular}{p{3.5cm}p{5.5cm}p{4.5cm}}
    \toprule
    Method & Mechanism & Representative Work \\
    \midrule
    FFT & Update all parameters; requires sufficient data to avoid overfitting & Standard TL; most SciML TL studies \\
    LFT / Layer Freezing & Freeze backbone; train only the last $N$ layers or task head & Early CV transfer~\cite{yosinski2014transferable}; NLP linear probing \\
    Adapter Modules & Insert small trainable MLP modules into frozen model & LLaMA-Adapter~\cite{zhang2023llama_adapter}; F-Adapter~\cite{zhang2026fadapter} \\
    LoRA & Inject trainable low-rank decomposition matrices into frozen weights & Hu et al.~\cite{hu2021lora}; current LLM fine-tuning standard \\
    \bottomrule
  \end{tabular}
\end{table}

\subsection{Prior Work by Transfer Domain}

Transfer learning has been explored across several engineering domains, though its adoption in SciML remains limited compared to computer vision and NLP.
We survey existing work by what is being transferred (physics and boundary conditions, simulation fidelity, or geometry) to identify the specific research gap this study addresses.

\paragraph{Transfer across physics and boundary conditions.}
The majority of SciML transfer learning studies focus on transferring across physics (e.g., varying PDE coefficients)~\cite{subramanian2023foundation} or boundary conditions (e.g., different inlet velocities or loading scenarios)~\cite{wang2025tlpinn}.
Chen et al.~\cite{chen2024pinn_thermal_jcise} demonstrated that physics-informed neural networks trained on randomly synthesized data can transfer to accelerate thermal simulations in additive manufacturing, effectively using synthetic data as a pretraining source.
Samuel and Ahmed~\cite{samuel2025continual} benchmarked continual learning strategies, a closely related paradigm, for 3D engineering regression problems, finding that naive fine-tuning leads to catastrophic forgetting of previously learned tasks, motivating the need for parameter-efficient approaches.
In these studies, the computational domain geometry remains fixed; only the governing equations, their coefficients, or the imposed conditions change.

\paragraph{Transfer across fidelity.}
A second class of transfer learning leverages data from cheaper, lower-fidelity simulations to bootstrap models for expensive high-fidelity targets.
Leng et al.~\cite{leng2024varfidelity} developed a variable-fidelity surrogate for aircraft multidisciplinary design optimization, fine-tuning a model trained on low-fidelity CFD data with a small number of high-fidelity samples.
Dang and Nguyen~\cite{dang2026deepoperator} applied deep operator learning for high-fidelity fluid flow field reconstruction from sparse sensor measurements, effectively transferring from a low-resolution representation to the full field.
Related but largely unexplored directions include transfer across turbulence modeling fidelity (e.g., Reynolds-Averaged Navier--Stokes (RANS) to Large Eddy Simulation (LES)) and across mesh resolutions, where the source and target share the same geometry and physics but differ in the level of modeling detail; these remain open challenges for future work.

\paragraph{Transfer across geometry.}
Transfer across geometry (adapting a model trained on one set of shapes to predict flow fields on a geometrically distinct shape) remains largely unexplored and constitutes the focus of the present study.
The few studies that address geometry do so in highly simplified settings: Wang et al.~\cite{wang2025tlpinn} compared FFT, lightweight fine-tuning, and LoRA for 2D PINNs with parametric domain boundary variations (circular to elliptical plate holes).
In all cases, the ``geometry transfer'' amounts to a smooth parametric deformation of the domain boundary within the same topological family, fundamentally different from the industrial scenario where each vehicle family has a distinct shape, different surface features, and non-overlapping geometric parameter spaces.
Chen et al.~\cite{chen2025geometric_repr} proposed self-supervised geometric representation learning for AI-driven surrogate modeling, demonstrating that learned geometric features can preserve fine-scale details across different shapes; however, their work focused on representation quality rather than transfer learning strategies, and operated at the academic benchmark scale.
More recently, Wu et al.~\cite{wu2026geopt} proposed GeoPT, which augments off-the-shelf geometries with synthetic dynamics to enable dynamics-aware self-supervised pre-training on over one million unlabeled samples; fine-tuning on real CFD data then yields labeled-data reductions of 20--60\% across automotive, aerospace, and marine benchmarks.
While GeoPT demonstrates the promise of geometric pre-training at scale, its data savings (20--60\%) remain moderate compared with the adapter-based strategy explored here (up to 70\% reduction), and the approach requires a separate synthetic-data generation pipeline that is itself computationally non-trivial.
Importantly, all existing PEFT applications in SciML, including LoRA on PINNs~\cite{wang2025tlpinn,jain2023hyperlora} and F-Adapter on neural operators~\cite{zhang2026fadapter}, operate on small networks ($\sim$10$^3$--10$^4$ parameters) in 2D settings; no prior work has applied these methods to a data-driven Transformer surrogate at industrial scale.

\paragraph{Domain consistency: an overlooked prerequisite.}
A fundamental distinction between prior PINN-based transfer learning and the industrial geometry-transfer setting concerns domain consistency between pretraining and fine-tuning.
In existing studies~\cite{wang2025tlpinn}, the computational domain is either fixed or undergoes smooth parametric deformation within a single topological family; consequently, the mapping between physical coordinates and the model's internal representation remains consistent across source and target tasks by construction.
In industrial vehicle aerodynamics, however, each vehicle family occupies a different spatial extent, features distinct surface topology, and spans an independent geometric parameter space; the source and target domains are fundamentally mismatched.
This mismatch introduces two requirements absent from same-domain settings: (1) the normalization mapping from physical space to the model's coordinate embedding must remain identical between pretraining and fine-tuning to preserve the geometric encoder's learned spatial relationships, and (2) the geometric encoder itself must be adapted, even partially, to accommodate shapes that lie outside its original training distribution.
To the authors' knowledge, no prior study has explicitly identified or addressed these domain-consistency requirements for geometry transfer in neural surrogates.

\subsection{Research Gap and Contributions}

Table~\ref{tab:gap_analysis} distills the preceding survey into a gap analysis.
Most SciML transfer learning studies use PINNs, with a smaller body of work on neural operators~\cite{zhang2026fadapter,dang2026deepoperator} or GNNs; transfer learning on Transformer-based surrogates, the architecture class that has driven the largest performance gains in NLP and vision, remains virtually absent.
Furthermore, existing studies operate almost exclusively at the academic benchmark scale: 2D domains, uniform grids of $64 \times 64$ to $256 \times 256$, and synthetically generated data; industrial 3D applications with $\sim$6\,million mesh points per case with complex geometry simulation data have not been addressed.
The present work occupies a unique position at the intersection of three underexplored dimensions: geometry transfer (rather than physics, boundary condition, or fidelity transfer), a systematic comparison of fine-tuning strategies (FFT vs.\ LFT vs.\ LoRA), and a Transformer-based surrogate architecture at industrial 3D scale.
To the best of our knowledge, no prior study has addressed any single combination of these dimensions, let alone all three simultaneously.

\begin{table}[ht]
  \centering
  \caption{Positioning of the present study within the SciML transfer learning landscape. \textbf{Bold} indicates this work.}
  \label{tab:gap_analysis}
  \small
  \begin{tabular}{lcc}
    \toprule
    Dimension & Prior Work & This Study \\
    \midrule
    Transfer target & Physics / BCs / fidelity & \textbf{Geometry (distinct families)} \\
    Fine-tuning method & FFT (default); LoRA on PINNs only & \textbf{FFT vs LFT vs LoRA} \\
    Architecture \& scale & PINN / FNO / GNN; academic 2D & \textbf{Transformer; industrial 3D ($\sim$6\,M pts)} \\
    \bottomrule
  \end{tabular}
\end{table}

\section{Methodology}\label{sec:methodology}

\subsection{Architecture and Dataset}\label{sec:architecture_dataset}

\begin{figure}[ht]
  \centering
  \includegraphics[width=0.95\textwidth]{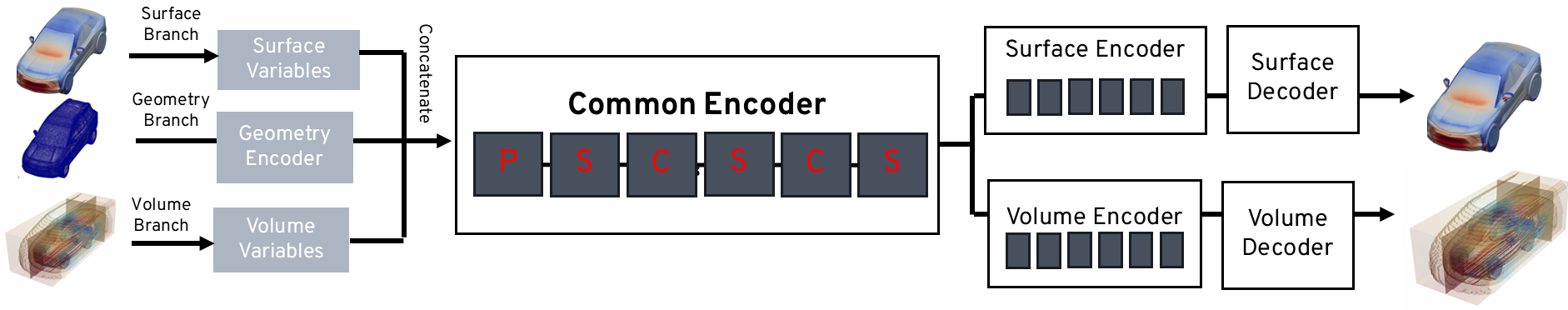}
  \caption{Schematic of the AB-UPT architecture reproduced from the original paper~\cite{alkin2025abupt}. Geometry supernodes, surface anchors, and volume anchors form three branches connected via self- and cross-attention. Note: this figure shows the original architecture for conceptual reference only; the model used in this study (Table~\ref{tab:model_config}) is deeper and wider, having been re-calibrated for production vehicle geometries.}
  \label{fig:abupt_arch}
\end{figure}

AB-UPT~\cite{alkin2025abupt} was adopted as the base architecture.
AB-UPT addresses the memory challenge of training on meshes with millions of points by employing a \emph{supernode} approach: instead of processing the entire mesh, the model trains on subsampled point sets drawn randomly at each epoch.
Geometry is encoded through \emph{supernodes}, while physical variables are sampled at \emph{anchor points} that are branched into surface and volume streams; self-attention and cross-attention among the branches ensure information flow across the entire domain while keeping memory consumption tractable.
Figure~\ref{fig:abupt_arch} reproduces the original architecture schematic from the AB-UPT paper~\cite{alkin2025abupt} for reference.
The model configuration actually used in this study (Table~\ref{tab:model_config}) differs substantially: the original architecture and hyperparameters, validated on the DrivAer benchmark~\cite{drivaerml}, did not meet the accuracy requirements for complex production vehicle geometries, and were therefore re-calibrated with increased model dimension and depth for greater expressivity.

Since the target engineering metric, aerodynamic drag, is computed entirely from surface quantities (pressure and wall shear stress integrated over the vehicle body), the model operates in surface-only mode in this study, predicting surface static pressure and surface friction force vectors at each query point.
Consequently, the common encoder is reduced to a single Perceiver block: cross-attention between surface and volume branches is unnecessary when only one branch is active, and the additional common encoder depth used in the original AB-UPT (which mediates surface--volume information exchange) provides no benefit in this configuration.

\begin{table}[ht]
  \centering
  \caption{Model configuration.}
  \label{tab:model_config}
  \small
  \begin{tabular}{ll}
    \toprule
    Parameter & Value \\
    \midrule
    Model dimension ($d_\text{model}$) & 512 \\
    Number of attention heads & 8 \\
    Geometry encoder depth & 6 Perceiver blocks~\cite{jaegle2021perceiver} (18.9\,M params, 30.8\%) \\
    Common encoder & One Perceiver \\
    Surface branch  & 12 Transformer blocks (38.8\,M params, 63.1\%) \\
    Surface decoder + bias & 3.8\,M params (6.1\%) \\
    Total parameters & 61.47\,M \\
    \bottomrule
  \end{tabular}
\end{table}

A key architectural feature relevant to this study is the explicit \emph{geometry encoder}, comprising 6 Perceiver blocks~\cite{jaegle2021perceiver} with 18.9\,million parameters (30.8\,\% of the total model).
This encoder transforms raw point cloud coordinates into a latent geometric representation that conditions all downstream predictions through cross-attention.
Whether this learned representation captures \emph{geometry-specific} features tied to the training families or \emph{general geometric} features (e.g., local curvature, relative spatial relationships) determines the feasibility of geometry transfer: if the encoder has overfit to the training geometries, transfer will fail regardless of the fine-tuning strategy; conversely, if it has learned generalizable geometric primitives, the latent representation should provide a useful starting point for adaptation.
The experiments in this study provide the first empirical test of this question.

\subsection{Dataset Description and Statistical Properties}\label{sec:dataset_characterization}
\begin{figure}[ht]
  \centering
  \includegraphics[width=0.7\textwidth]{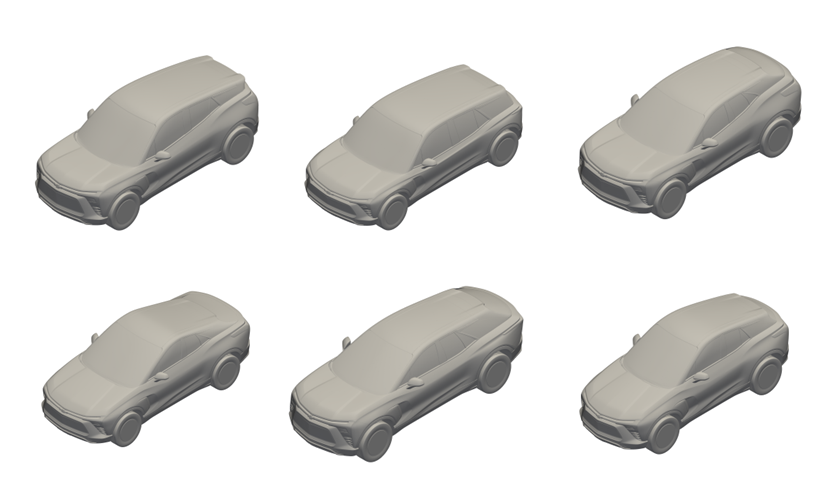}
  \caption{Sample geometric variations of the Mid-size SUV family. Ten geometric parameters are varied at 10 levels each, yielding $\sim$100 configurations per family.}
  \label{fig:blazer_variations}
\end{figure}

A vehicle external aerodynamics dataset comprising 511 cases was prepared across five geometrically distinct vehicle families spanning a broad range of sizes: a Small Crossover, a Mid-size Sport Utility Vehicle (SUV), and three Large-size SUVs (\#1, \#2, and \#3).
These families differ not merely in parametric dimensions but in fundamental shape topology: distinct roof profiles, underbody architectures, A-pillar angles, and rear-end geometries that cannot be represented as smooth deformations of a single base shape.
For each family, the geometry was varied through 10 geometric parameters (including wheelbase, roof height, front windshield angle, and others), each discretized into 10 levels, yielding approximately 100 geometric configurations per family via Latin Hypercube Sampling~\cite{mckay1979lhs}.
CFD simulations were conducted for all 511 geometries using the commercial Lattice Boltzmann solver PowerFLOW~\cite{powerflow}; transient simulations were performed and the converged solutions were time-averaged during postprocessing. Other than the geometric variations, all simulation settings were held constant across the dataset, including boundary conditions, meshing scheme, and solver parameters. Sample geometric variations of the Mid-size SUV family are shown in Figure~\ref{fig:blazer_variations}, and geometric bounding-box extents are listed in Table~\ref{tab:geometry_bounds}.

\begin{figure}[ht]
  \centering
  \includegraphics[width=0.7\textwidth]{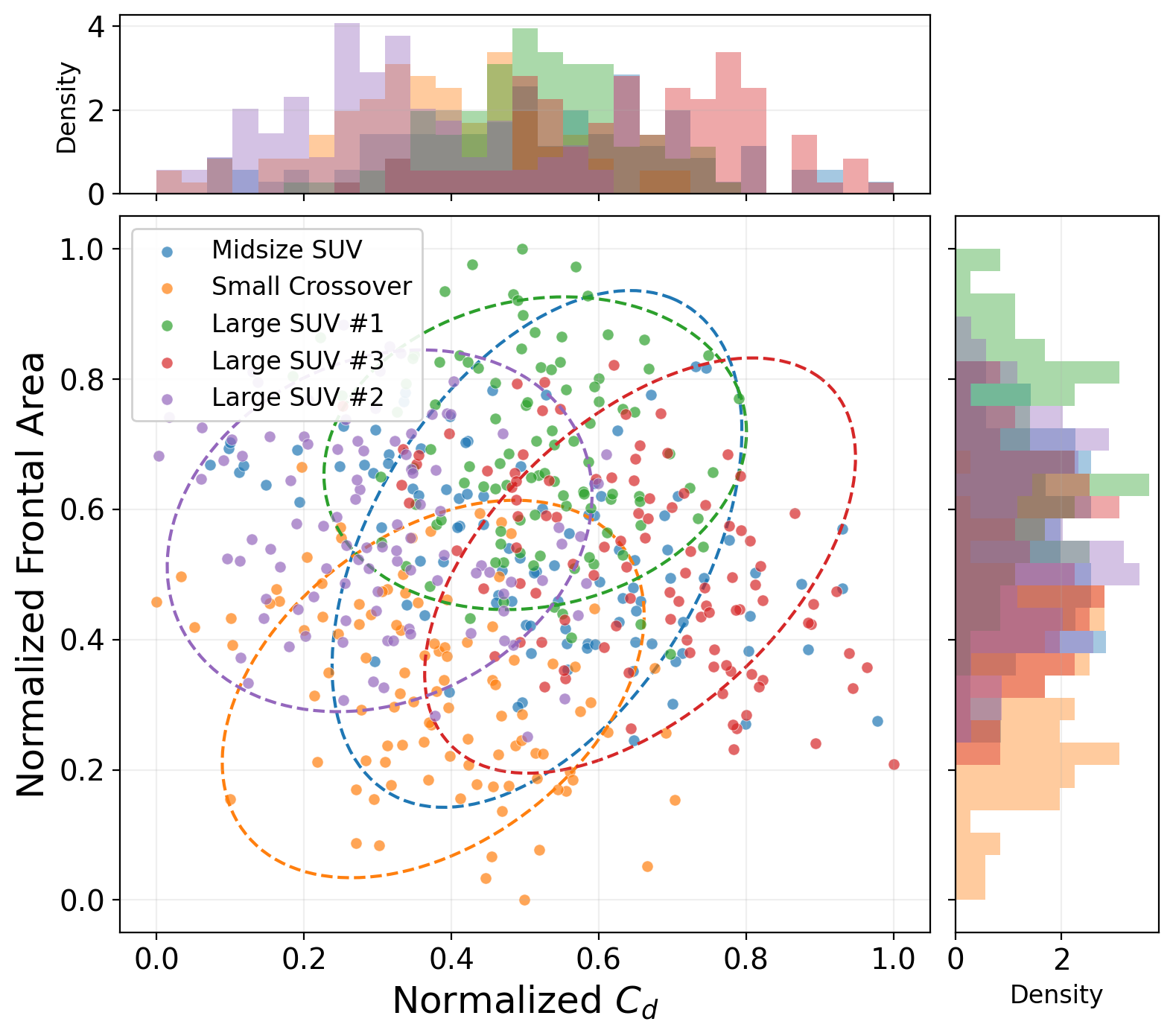}
  \caption{Normalized frontal area vs.\ $C_d$ distribution for five vehicle families. Dashed ellipses indicate approximate 2$\sigma$ contours.}
  \label{fig:area_cd}
\end{figure}

\begin{table}[ht]
  \centering
  \caption{Geometric bounding-box extents (meters) for the five vehicle families. \textbf{Bold} = column maximum; \textit{italic} = column minimum.}
  \label{tab:geometry_bounds}
  \small
  \begin{tabular}{c cc cc cc}
    \toprule
    \multirow{2}{*}{Vehicle} & \multicolumn{2}{c}{X (m)} & \multicolumn{2}{c}{Y (m)} & \multicolumn{2}{c}{Z (m)} \\
    \cmidrule(lr){2-3} \cmidrule(lr){4-5} \cmidrule(lr){6-7}
     & Min & Max & Min & Max & Min & Max \\
    \midrule
    Small Crossover   & $\mathbf{-2.53}$ & $-1.20$ & 0 & $\mathit{2.37}$ & 1.20 & $\mathit{1.85}$ \\
    Mid-size SUV      & $-2.72$ & $-1.26$ & 0 & 2.83 & 1.25 & 1.91 \\
    Large SUV \#1     & $\mathit{-2.85}$ & $\mathit{-1.30}$ & 0 & $\mathbf{3.04}$ & $\mathbf{1.29}$ & $\mathbf{2.01}$ \\
    Large SUV \#2     & $-2.81$ & $-1.25$ & 0 & 3.02 & 1.24 & 1.99 \\
    Large SUV \#3     & $-2.60$ & $\mathbf{-1.19}$ & 0 & 2.67 & $\mathit{1.19}$ & 1.97 \\
    \bottomrule
  \end{tabular}
\end{table}
The normalized frontal area vs.\ drag coefficient distributions are shown in Figure~\ref{fig:area_cd}.
A critical observation is that, in the aerodynamic feature space, no single family constitutes a completely out-of-distribution (OOD) case relative to the others.
The bounding-box extents in Table~\ref{tab:geometry_bounds} show that the families span overlapping size ranges: the three Large SUVs share similar longitudinal and lateral dimensions, the Mid-size SUV overlaps with the upper range of the Small Crossover and the lower range of the Large SUVs, and even the most compact family (Small Crossover) shares its Z-range with the lower bound of the Large SUVs.
In the $C_d$--frontal-area plane, every family's distribution partially intersects with at least one other, meaning that for any held-out geometry the pretrained model has already seen aerodynamically similar configurations, albeit realized by topologically different shapes.
This partial overlap is precisely what makes transfer learning feasible: the pretrained encoder is never asked to extrapolate into a completely unseen aerodynamic regime, but rather to generalize its learned geometric representations to new shape topologies that produce similar flow physics.
At the same time, the distinct clustering confirms that the families are not trivially interchangeable: a model trained on four families cannot simply interpolate to predict the fifth without adaptation, because the geometric features (e.g., roof curvature, underbody topology, A-pillar angle) that drive aerodynamic differences are family-specific even when the integrated forces overlap.

This dataset was chosen over publicly available alternatives such as DrivAerML~\cite{drivaerml} because open-source databases are typically based on sedan-type geometries and do not cover the SUV segments that dominate current sales volumes; internal evaluations confirmed that published hyperparameter settings developed on these benchmarks do not generalize well to production-level SUV geometries.
Most importantly, the dataset provides five topologically distinct vehicle families, enabling leave-one-family-out evaluation of geometry transfer, whereas open-source databases typically offer parametric variations of a single base geometry.
The dataset is available to researchers upon request and approval from General Motors; interested parties may contact the corresponding author for access.
Model performance is evaluated using aerodynamic drag forces computed by surface integration; raw force values in Newtons are used rather than dimensionless coefficients ($C_d$, $C_l$), which depend on the choice of reference area and pressure that can be defined differently across vehicle programs.

\subsection{Transfer Learning Strategies}\label{sec:finetuning_strategies}

All three transfer learning strategies evaluated in this study start from the same pretrained checkpoint and use the same fine-tuning dataset from the held-out vehicle family, with training hyperparameters held constant across strategies to ensure a fair comparison.
Using identical hyperparameters is a deliberate design choice: it isolates the effect of the fine-tuning \emph{mechanism} (which layers are updated, how updates are constrained) from confounding hyperparameter interactions.
The shared learning rate ($1 \times 10^{-5}$) was already reduced to one-third of the pretraining rate, itself a standard FFT best practice, and weight decay and gradient clipping were applied uniformly.

\paragraph{Fine-tuning dataset.}
For each leave-one-family-out experiment, approximately 100 CFD simulations are available for the held-out target family.
Of these, only 30 cases (approximately 30\,\% of the available per-family data) are randomly sampled for fine-tuning: 20 are used for training and 10 for validation, yielding a total of 30 target-family samples consumed by the adaptation process.
The remaining $\sim$70 cases constitute an unseen test set for evaluation.
This deliberate restriction to 30 samples reflects the practical constraint that, at the onset of a new vehicle program, only a limited number of CFD runs are typically available.
A 5-fold cross-validation study confirms that LoRA performance is insensitive to the specific random selection of these 30 samples; details are presented in Section~\ref{sec:lora_results}.

\paragraph{FFT.}
All 61.47\,million parameters of the pretrained model are updated using the target family data.
This is the standard transfer learning approach and serves as the primary baseline; the pretrained normalization statistics are preserved to maintain consistent input scaling.
FFT represents the maximum-freedom end of the design space: every layer is free to change at full rank.
The primary risk of FFT in the low-data regime is catastrophic forgetting, where the model overwrites useful pretrained representations while overfitting to the small target set.
Common mitigation strategies include reducing the learning rate relative to pretraining, applying aggressive weight decay or early stopping, using elastic weight consolidation (EWC)~\cite{kirkpatrick2017ewc} to penalize deviations from pretrained weights, or replaying a subset of the pretraining data during fine-tuning~\cite{rebuffi2017icarl}.
In this study, we adopt the first two (the fine-tuning learning rate is set to one-third of the pretraining rate ($1 \times 10^{-5}$ vs.\ $3 \times 10^{-5}$) and the best model checkpoint (lowest validation loss across all epochs) is selected for evaluation, providing an implicit early-stopping effect) but do not employ EWC or data replay, as these require access to the pretraining data distribution which may not be available in a deployment scenario where the pretrained checkpoint is shared across organizational boundaries.

\paragraph{LFT.}
LFT freezes the geometry encoder (6 blocks, 18.9\,million parameters), the encoder projection layers, the first 10 of 12 surface blocks, and all position-related embeddings~\cite{yosinski2014transferable}.
Only the last two surface blocks, the surface decoder, and the surface output bias remain trainable, totaling 6.83\,million parameters (11.1\,\% of the model).
This represents the classical transfer learning paradigm (preserve the pretrained feature extractor and adapt only the task-specific head and final representation layers~\cite{zhuang2020comprehensive}) and has been the dominant approach in computer vision transfer where early CNN layers learn universal edge and texture features.
Critically, in LFT the geometry encoder is entirely frozen; if the encoder's learned representation is not sufficiently general for the new geometry, the downstream layers cannot compensate.

\paragraph{LoRA.}
LoRA~\cite{hu2021lora} addresses the limitations of both preceding strategies.
FFT updates every element of a $d \times d$ weight matrix at each layer, requiring $\mathcal{O}(d^2)$ trainable parameters per layer, prohibitively expensive for large models and prone to overfitting when data is scarce.
LFT avoids this cost by restricting updates to a subset of layers, but at the price of reduced representational capacity in the frozen layers.
LoRA resolves this trade-off by keeping the original pretrained weights $W_0$ entirely frozen, thereby preserving the learned representations, and injecting two small trainable matrices of rank $r \ll d$ that together parameterize a low-rank weight update $\Delta W = BA$ with only $2 \times r \times d$ parameters per layer instead of $d^2$.
This yields an effective update at every layer (maintaining representational reach across the full network) while constraining each layer's modification to a low-dimensional manifold that acts as an implicit regularizer in the low-data regime.
Formally, the forward pass becomes:
\begin{equation}\label{eq:lora}
  h = W_0 x + \frac{\alpha}{r}\, B A\, x
\end{equation}
where $A \in \mathbb{R}^{r \times d}$ and $B \in \mathbb{R}^{d \times r}$ are low-rank factors, $r$ is the rank, and $\alpha$ is a scaling factor that controls the magnitude of the adaptation relative to the pretrained weights.
Adapters are injected into all linear layers of the model (query, key, and value projections, attention output projections, and both MLP layers) across the geometry encoder, surface branch, and shared blocks, yielding 158 adapted layers with 10.13\,million trainable parameters (16.5\,\%).
The key hyperparameters are $r = 64$ and $\alpha = 128$ ($= 2r$), selected based on a rank sweep presented in Section~\ref{sec:lora_results}; the implementation uses the PEFT library~\cite{peft2022}.
Unlike LFT, LoRA adapts \emph{all} layers including the geometry encoder, but constrains each layer's update to a rank-64 manifold rather than granting full freedom.

The conceptual distinction between LFT and LoRA is critical to the experimental design.
LFT grants \emph{full freedom} to a \emph{subset of layers}, specifically the last two blocks, while keeping the geometry encoder entirely frozen; LoRA grants \emph{constrained freedom} to \emph{all layers} including the geometry encoder.
Comparing the two therefore disentangles \emph{where} to adapt (layer selection, as in LFT) from \emph{how much} to adapt (rank constraint, as in LoRA), and reveals which mechanism is responsible for successful geometry transfer in the low-data regime.
Table~\ref{tab:strategy_comparison} summarizes both the architectural differences and the complete experimental matrix.

\begin{table}[ht]
  \centering
  \caption{Experimental matrix and strategy comparison. Fine-tuning rows are repeated for all five leave-one-out configurations (15 runs total).}
  \label{tab:strategy_comparison}
  \small
  \begin{tabular}{lcccc}
    \toprule
    Experiment & Base Model & Training Data & Trainable Params & Geometry Encoder \\
    \midrule
    Pretrain         & --          & 4 fam.\ ($\sim$410) & 61.47\,M (100\%)  & Trained         \\
    From-scratch (103) & --       & 5 fam.\ (103)& 61.47\,M (100\%)  & Trained         \\
    From-scratch (30)  & --       & 5 fam.\ (30) & 61.47\,M (100\%)  & Trained         \\
    \midrule
    FFT              & Pretrain     & 20             & 61.47\,M (100\%)  & Updated (full)  \\
    LFT              & Pretrain     & 20             & 6.83\,M (11.1\%)  & {Frozen} \\
    {LoRA}    & Pretrain     & 20  & {10.13\,M (16.5\%)} & Adapted (rank-64) \\
    \bottomrule
  \end{tabular}
\end{table}

\paragraph{Data normalization.}
All input and output variables undergo normalization prior to model training.
Physical field variables (surface pressure, friction force vectors) are standardized using z-score normalization: $\hat{x} = (x - \mu) / \sigma$, where $\mu$ and $\sigma$ are the per-component mean and standard deviation computed across the entire pretraining dataset.
Spatial coordinates are normalized via min-max scaling to a fixed range $[0, 1000]$: $\hat{p} = (p - p_{\min}) / (p_{\max} - p_{\min}) \times 1000$, where $p_{\min}$ and $p_{\max}$ are the coordinate extrema observed across all pretraining geometries.
Crucially, these normalization statistics, both the field moments and the coordinate bounds, \textbf{must} be frozen at pretraining time and carried over unchanged to all subsequent fine-tuning and inference stages; recomputing them from the target data alone would destroy the pretrained representations (see Section~\ref{sec:discussion} for an empirical observation of this failure mode).
This strict requirement ensures that the positional encoding space seen by the geometry encoder remains identical regardless of whether the input geometry is from the original training distribution or a new target family, preserving the learned spatial relationships within the pretrained representations.

\subsection{Experimental Design}\label{sec:experimental_design}

For each of the five vehicle families, a \emph{leave-one-family-out} pretrained model was prepared: four families were used for pretraining (approximately 410 cases, 80/10/10 train/val/test split) and the fifth was held out entirely.
Pretraining used a learning rate of $3 \times 10^{-5}$ with cosine annealing~\cite{loshchilov2017sgdr} over 500 epochs on 4$\times$ NVIDIA A100 Graphics Processing Units (GPUs).
Each fine-tuning experiment then used 20 training samples and 10 validation samples from the held-out family, with the remaining $\sim$70 cases reserved as an unseen test set.
All three strategies shared the hyperparameters listed in Table~\ref{tab:ft_hyperparams}; the pretrain configuration is shown alongside for reference.

\begin{table}[ht]
  \centering
  \caption{Training hyperparameters for pretraining and fine-tuning.}
  \label{tab:ft_hyperparams}
  \small
  \begin{tabular}{lcc}
    \toprule
    Hyperparameter & Pretrain & Fine-tuning \\
    \midrule
    Learning rate & $3 \times 10^{-5}$ & $1 \times 10^{-5}$ \\
    Scheduler & Cosine annealing & Cosine annealing \\
    Warmup epochs & 10 & 3 \\
    Total epochs & 500 & 50 \\
    Weight decay & $1 \times 10^{-5}$ & $1 \times 10^{-5}$ \\
    Gradient clipping & 3.0 & 3.0 \\
    GPUs & 4$\times$ A100 (80\,GB) & 4$\times$ A100 (80\,GB) \\
    \bottomrule
  \end{tabular}
\end{table}

\subsection{Evaluation Metrics}\label{sec:evaluation_metrics}

Model performance is assessed at two levels: integrated aerodynamic forces (the engineering deliverable) and pointwise field accuracy (diagnostic of spatial prediction quality).

\paragraph{Force metrics.}
For each method and held-out family, the model predicts surface pressure and friction fields at all mesh points; these are integrated over the vehicle surface to yield scalar drag force values (in Newtons) for each test case.
Three metrics are reported on these integrated forces:
\begin{itemize}
  \item \textbf{Coefficient of determination} ($R^2$): computed separately for pressure-drag and shear-drag components across the held-out test set ($\sim$70 cases per family). $R^2 = 1$ indicates perfect prediction, $R^2 = 0$ equals predicting the mean, and $R^2 < 0$ indicates the model is worse than a constant-mean predictor.
  \item \textbf{Mean Absolute Error} (MAE): average absolute deviation between predicted and ground-truth force values, reported in Newtons.
  \item \textbf{RMSE}: root-mean-square deviation between predicted and ground-truth force values, reported in Newtons.
\end{itemize}

\paragraph{Field metrics.}
Pointwise prediction quality on the surface mesh is quantified by two complementary metrics, computed separately for pressure and friction fields:
\begin{itemize}
  \item \textbf{Relative $L_2$ error} (rel.\ $L_2$): defined as $\|y_{\text{pred}} - y_{\text{true}}\|_2 \;/\; \|y_{\text{true}}\|_2$, where norms are taken over all mesh points for a given test case. For surface pressure, values are reported in units of $\times 10^{-4}$ for readability.
  \item \textbf{Normalized RMSE} (NRMSE): the pointwise RMSE divided by the range of the ground-truth field, expressed as a percentage: $\text{NRMSE} = \text{RMSE}_{\text{field}} / (y_{\max} - y_{\min}) \times 100\,\%$.
\end{itemize}
All metrics are computed per test case and then averaged across the held-out test set; tables report mean $\pm$ standard deviation across the five leave-one-out experiments.

\section{Results}\label{sec:results}

\subsection{Pretrained Model Performance}\label{sec:pretrain_performance}

\begin{figure}[h!]
  \centering
  \includegraphics[width=0.85\textwidth]{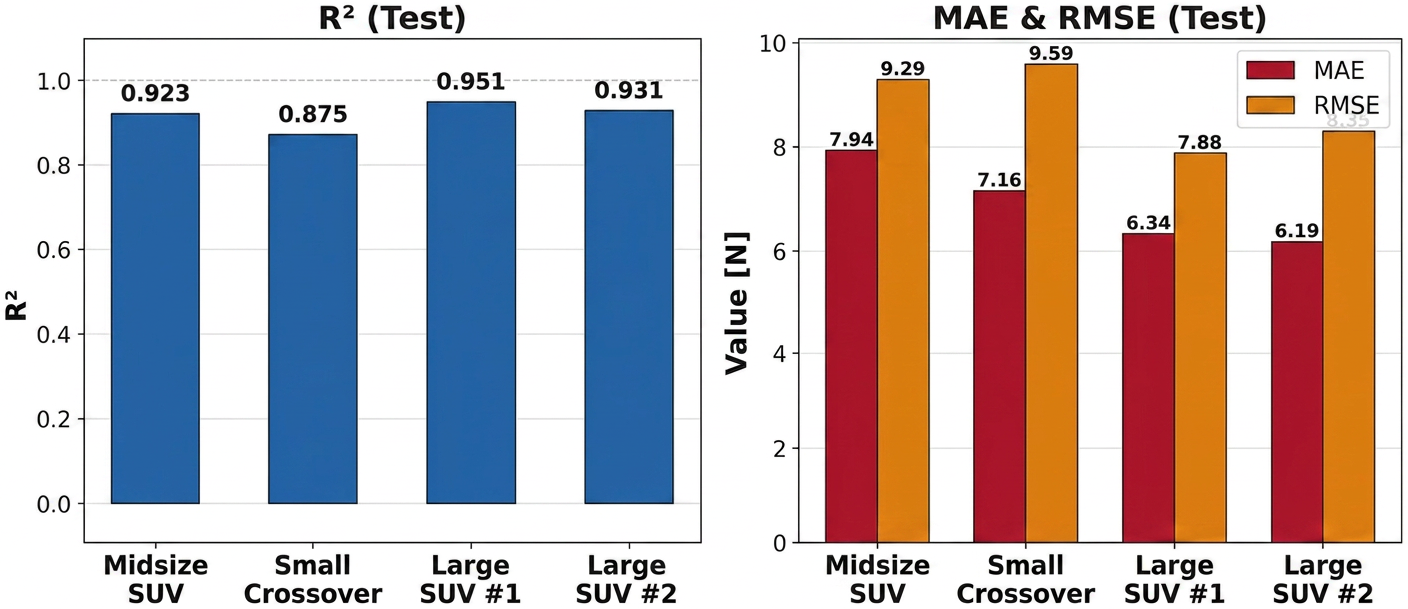}
  \caption{In-distribution performance of the pretrained model on the four training families.  Left: pressure-drag $R^2$ on held-out test splits.  Right: MAE and RMSE of predicted drag force.}
  \label{fig:pretrain_indist}
\end{figure}

Before examining the transfer learning strategies, we establish the pretrained model's baseline capabilities.
For each leave-one-out experiment, the model was pretrained on four of the five vehicle families (e.g., Mid-size SUV, Small Crossover, Large SUV~\#1, and Large SUV~\#2 when Large SUV~\#3 is held out; 411 samples in each case) and evaluated on the held-out test splits.
We describe the Large SUV~\#3 held-out case as a representative example; the remaining four experiments follow the same protocol.
Figure~\ref{fig:pretrain_indist} summarizes the in-distribution performance: the pretrained model achieves $R^2 > 0.87$ for integrated aerodynamic drag across all four families, with MAE consistently below 8\,N (approximately 2\,\% of total drag).
These results confirm that the pretrained model has learned robust representations of the pressure and friction fields for the geometries it has seen.

\begin{figure}[ht]
  \centering
  \begin{subfigure}[b]{0.32\textwidth}
    \includegraphics[width=\textwidth]{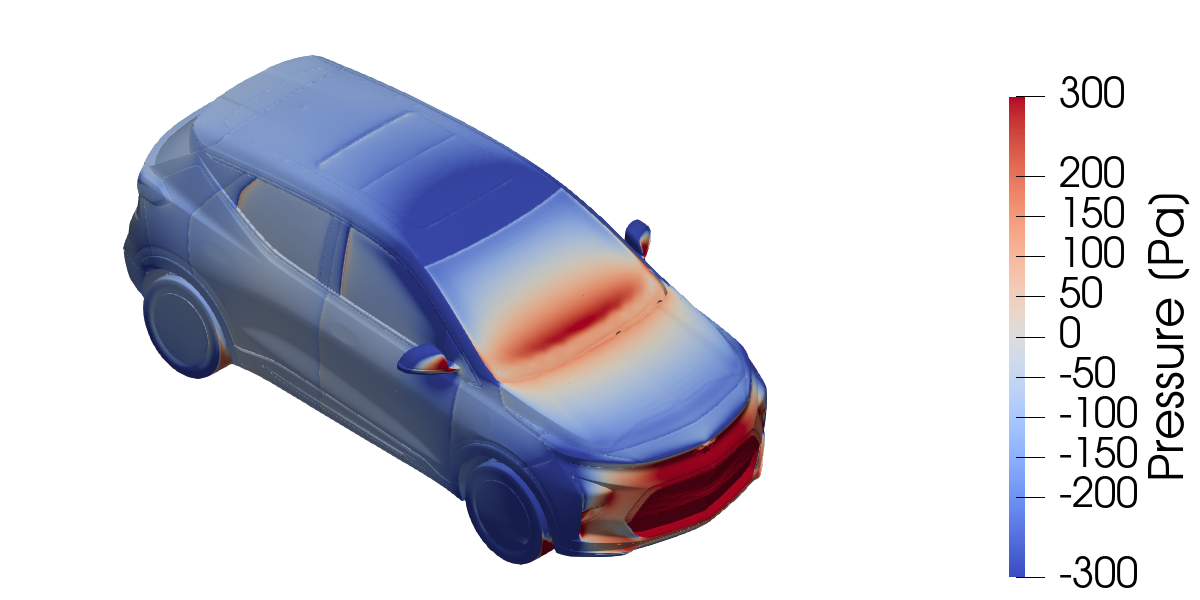}
    \caption{Small Crossover: Ground truth}
  \end{subfigure}\hfill
  \begin{subfigure}[b]{0.32\textwidth}
    \includegraphics[width=\textwidth]{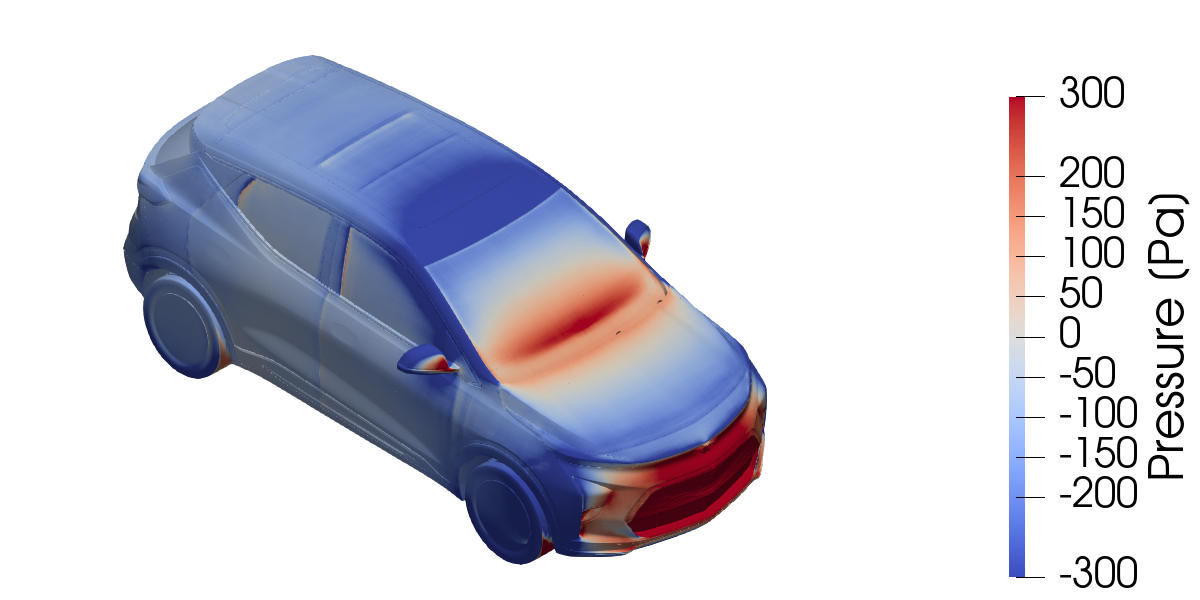}
    \caption{Small Crossover: Prediction}
  \end{subfigure}\hfill
  \begin{subfigure}[b]{0.32\textwidth}
    \includegraphics[width=\textwidth]{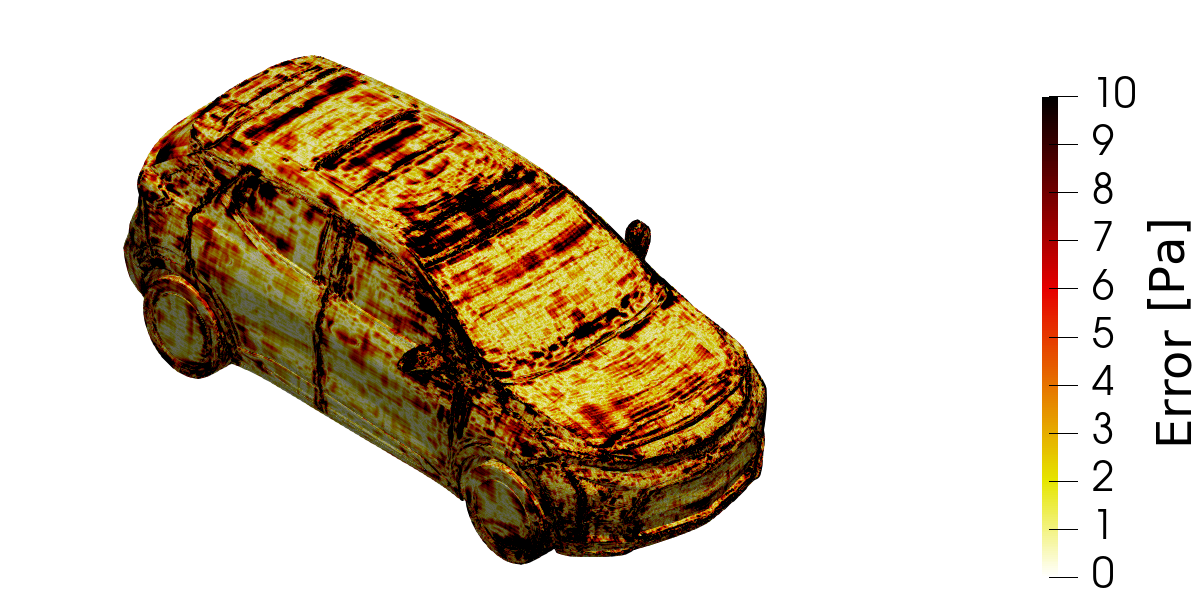}
    \caption{Small Crossover: $|$Error$|$}
  \end{subfigure}\\[4pt]
  \begin{subfigure}[b]{0.32\textwidth}
    \includegraphics[width=\textwidth]{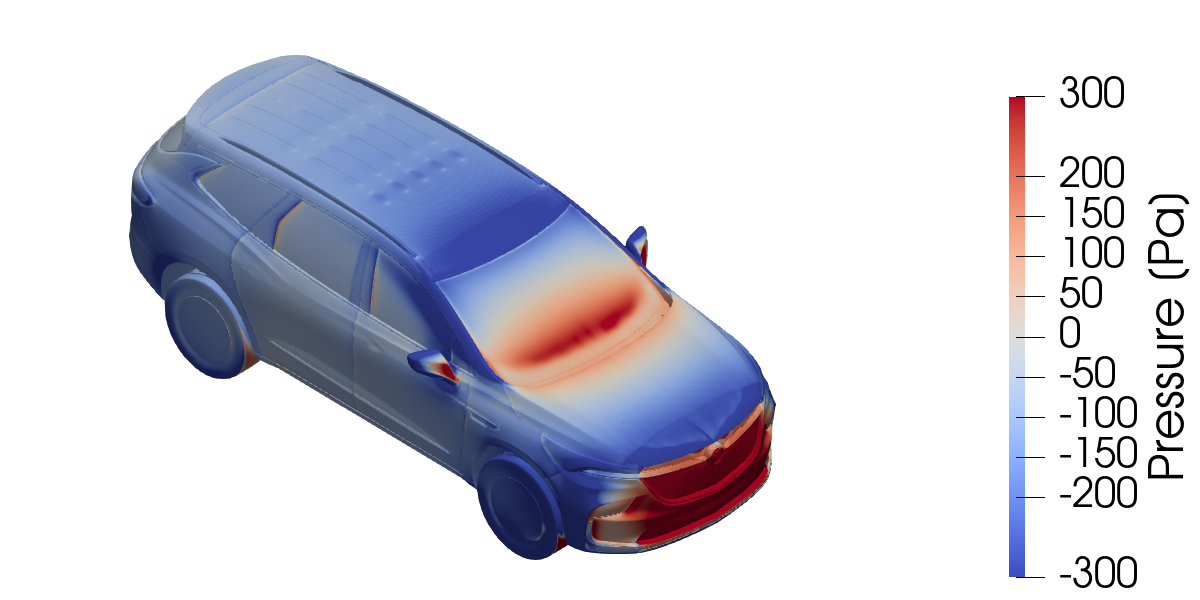}
    \caption{Large SUV \#1: Ground truth}
  \end{subfigure}\hfill
  \begin{subfigure}[b]{0.32\textwidth}
    \includegraphics[width=\textwidth]{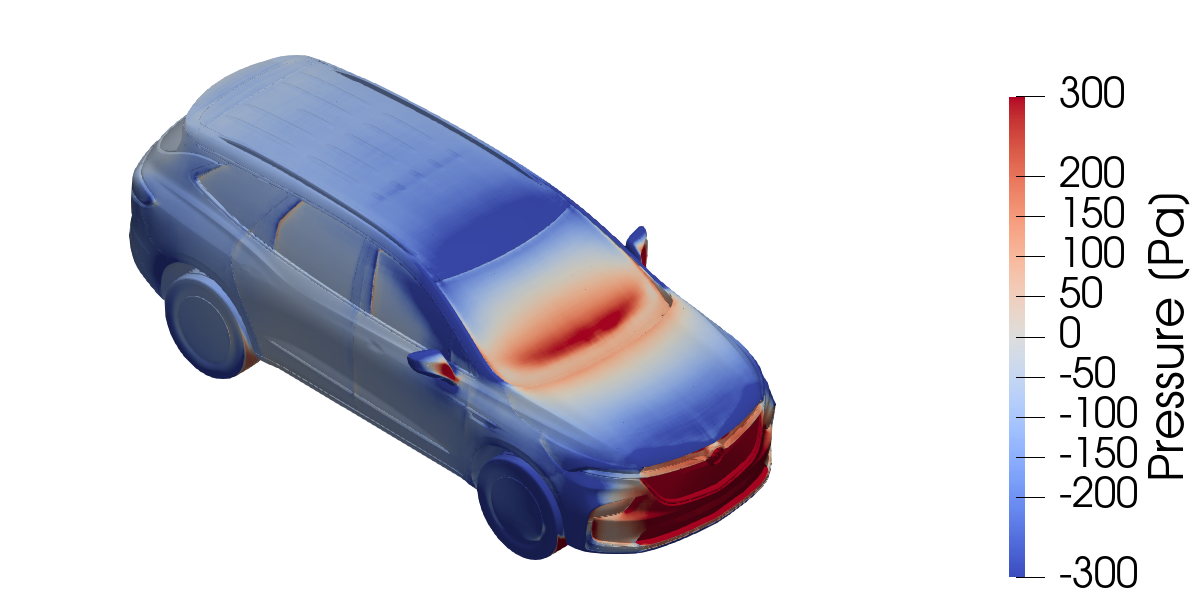}
    \caption{Large SUV \#1: Prediction}
  \end{subfigure}\hfill
  \begin{subfigure}[b]{0.32\textwidth}
    \includegraphics[width=\textwidth]{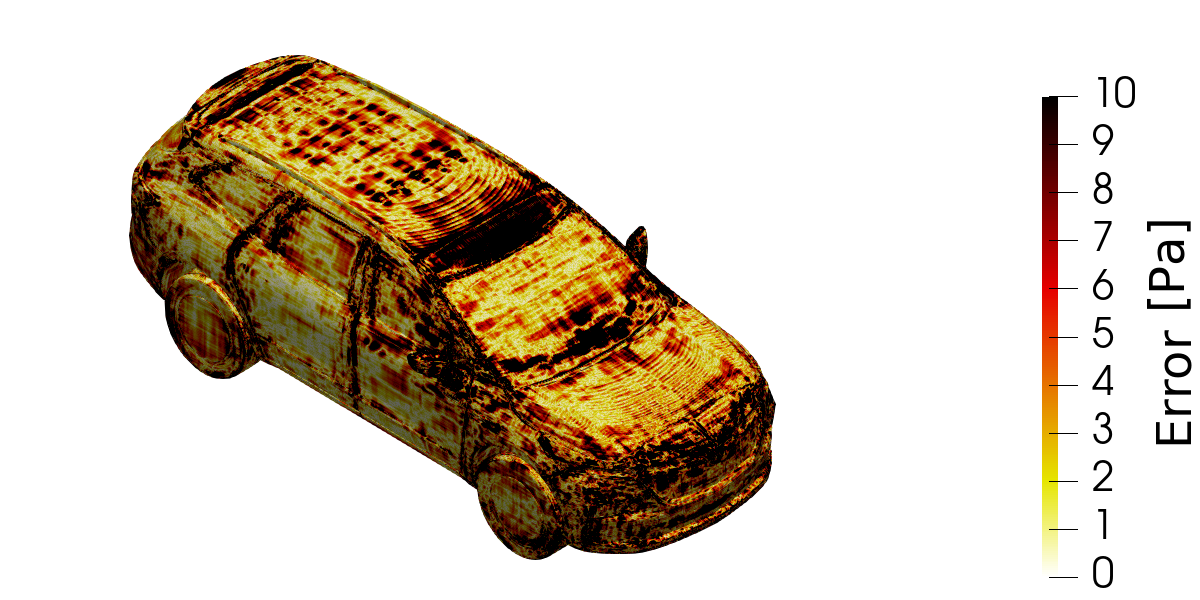}
    \caption{Large SUV \#1: $|$Error$|$}
  \end{subfigure}
  \caption{Surface pressure predictions of the pretrained model (leave-Large-SUV-\#3-out checkpoint; both vehicles are \emph{in-distribution}) on baseline configurations. Top: Small Crossover (RMSE~$= 19.5$\,Pa, drag error $= 0.43$\,N). Bottom: Large SUV \#1 (RMSE~$= 21.5$\,Pa, drag error $= 3.9$\,N). Color range: $[-300, 300]$\,Pa (pressure), $[0, 10]$\,Pa (error). Errors concentrate near geometric discontinuities, confirming the model captures smooth pressure topology with high fidelity.}
  \label{fig:pressure_pretrain}
\end{figure}

Figure~\ref{fig:pressure_pretrain} provides qualitative evidence of the pretrained model's surface-level prediction fidelity on representative baseline configurations from two distinct vehicle families.
Importantly, both vehicles shown are \emph{in-distribution}: the predictions are produced by the leave-Large-SUV-\#3-out checkpoint, which was trained on all four non-Large-SUV-\#3 families including both Small Crossover and Large SUV~\#1.
For Small Crossover, the pretrained model achieves a pressure RMSE of 19.5\,Pa (NRMSE~$= 0.68$\,\%, field $R^2 = 0.992$) and a drag error of only 0.43\,N (0.10\,\%); for Large SUV \#1, RMSE~$= 21.5$\,Pa ($R^2 = 0.987$, drag error 3.9\,N).
Errors are concentrated near geometric discontinuities (A-pillar, wheel arches, rear spoiler) rather than distributed uniformly, indicating that the model's learned representations faithfully capture the smooth pressure topology.
The high fidelity on these in-distribution cases motivates the question of whether such representational quality transfers to geometries unseen during training.

\begin{figure}[h!]
  \centering
  \includegraphics[width=0.52\textwidth]{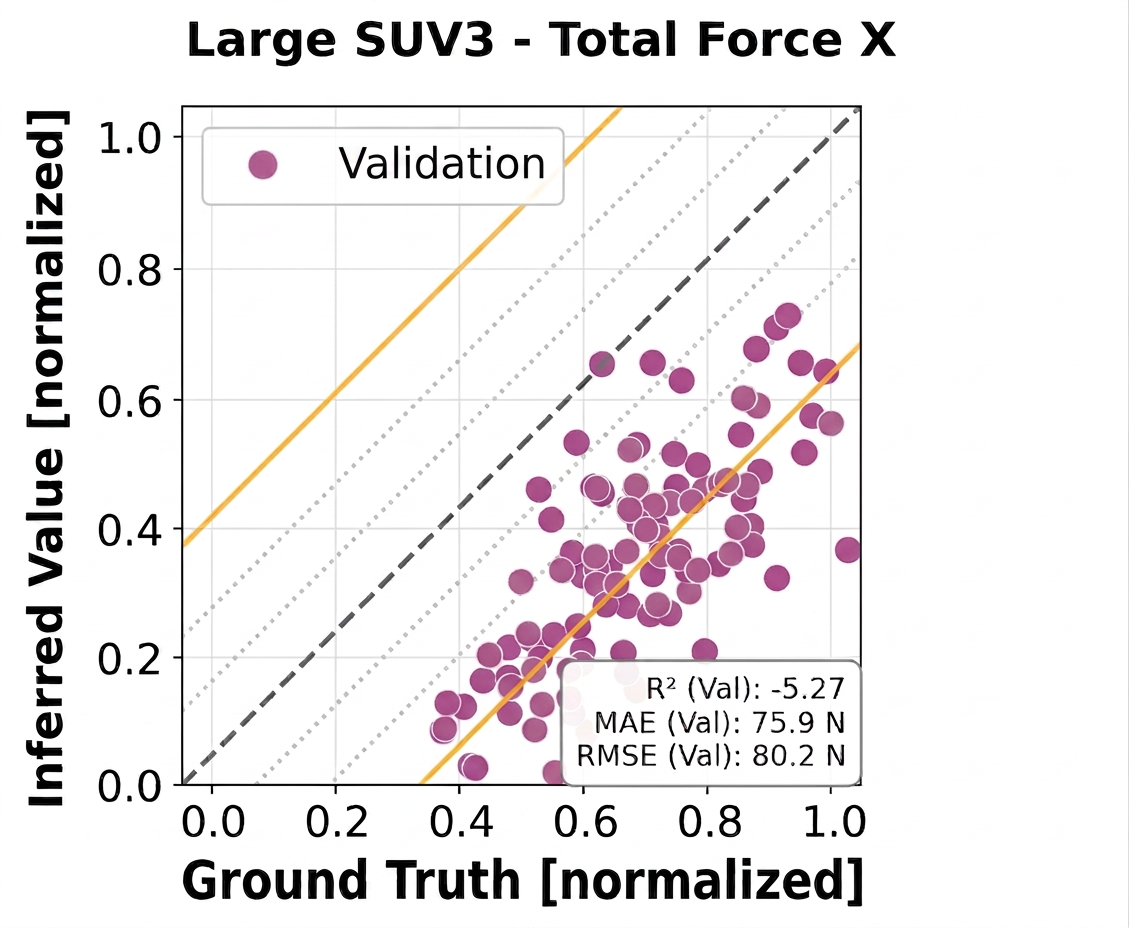}
  \caption{Zero-shot inference on the held-out Large SUV~\#3 family using the pretrained model without any fine-tuning.  Despite a weak positive correlation indicating partial physics understanding, predictions exhibit strong bias and large scatter ($R^2 = -5.27$, MAE~$= 75.9$\,N).}
  \label{fig:zeroshot_k3}
\end{figure}

A natural question is whether this strong in-distribution performance generalizes to unseen vehicle families.
To answer this, the pretrained model is applied directly to each held-out vehicle family without any parameter updates, a zero-shot evaluation.
Figure~\ref{fig:zeroshot_k3} shows a representative case (Large SUV~\#3): the predicted drag forces correlate weakly with the ground truth but suffer from a strong systematic bias and large scatter ($R^2 = -5.27$, MAE~$= 75.9$\,N).
Quantitatively, the zero-shot evaluation across all five held-out families yields a mean surface-pressure relative $L_2$ error of $14.08 \times 10^{-4}$ (NRMSE~$= 4.00$\,\%) and a friction relative $L_2$ of 0.813 (NRMSE~$= 3.06$\,\%), approximately $4.4\times$ and $4.1\times$ worse than the best adapted model (LoRA, Section~\ref{sec:lora_results}), respectively.
The fact that some correlation persists indicates that the pretrained encoder has learned partially transferable representations of the underlying flow physics; however, the magnitude of the errors renders zero-shot predictions unusable for engineering design.
This gap motivates the systematic evaluation of adaptation strategies presented in the following subsections.

\subsection{Full Fine-Tuning}\label{sec:fft_results}

FFT updates all 61.47\,million parameters starting from the pretrained checkpoint.
Table~\ref{tab:fft_results} presents the per-vehicle results.
The defining characteristic is extreme vehicle-dependent variability: pressure-drag $R^2$ ranges from $0.70$ (Large SUV \#1) to $-0.17$ (Small Crossover), with a standard deviation of 0.33 across families.
This instability arises because 61.47\,million parameters trained on only 20 samples yield a ratio of over 3 million parameters per sample, placing the optimizer in an extremely ill-conditioned regime.
Whether the optimizer converges to a reasonable basin depends on the specific target geometry, a factor that cannot be predicted in advance, making FFT fundamentally unreliable for deployment.

\begin{table}[ht]
  \centering
  \caption{FFT test-set results across five leave-one-out experiments. Force $R^2$ is computed from integrated surface forces; field metrics are pointwise surface errors.}
  \label{tab:fft_results}
  \footnotesize
  \begin{tabular}{lcc cc cc}
    \toprule
    & \multicolumn{2}{c}{Force $R^2$} & \multicolumn{2}{c}{Surface Pressure} & \multicolumn{2}{c}{Surface Friction} \\
    \cmidrule(lr){2-3} \cmidrule(lr){4-5} \cmidrule(lr){6-7}
    Held-out Vehicle & Pressure & Shear & rel.\ $L_2$ ($\times 10^{-4}$) & NRMSE (\%) & rel.\ $L_2$ & NRMSE (\%) \\
    \midrule
    Small Crossover   & $-0.165$ & $0.072$ & $5.14$ & $1.88$ & $0.402$ & $1.23$ \\
    Mid-size SUV      & $0.228$  & $0.930$ & $3.92$ & $1.21$ & $0.262$ & $1.46$ \\
    Large SUV \#1     & $0.702$  & $0.716$ & $4.08$ & $1.10$ & $0.231$ & $0.70$ \\
    Large SUV \#2     & $0.645$  & $0.950$ & $4.81$ & $0.49$ & $0.400$ & $0.51$ \\
    Large SUV \#3     & $0.588$  & $0.902$ & $4.44$ & $1.42$ & $0.298$ & $1.82$ \\
    \midrule
    \textbf{Mean $\pm$ Std} & $\mathbf{0.400 \pm 0.33}$ & $\mathbf{0.714 \pm 0.33}$ & $\mathbf{4.48 \pm 0.45}$ & $\mathbf{1.22 \pm 0.45}$ & $\mathbf{0.319 \pm 0.07}$ & $\mathbf{1.14 \pm 0.48}$ \\
    \bottomrule
  \end{tabular}
\end{table}

\begin{figure}[ht]
  \centering
  \includegraphics[width=\textwidth]{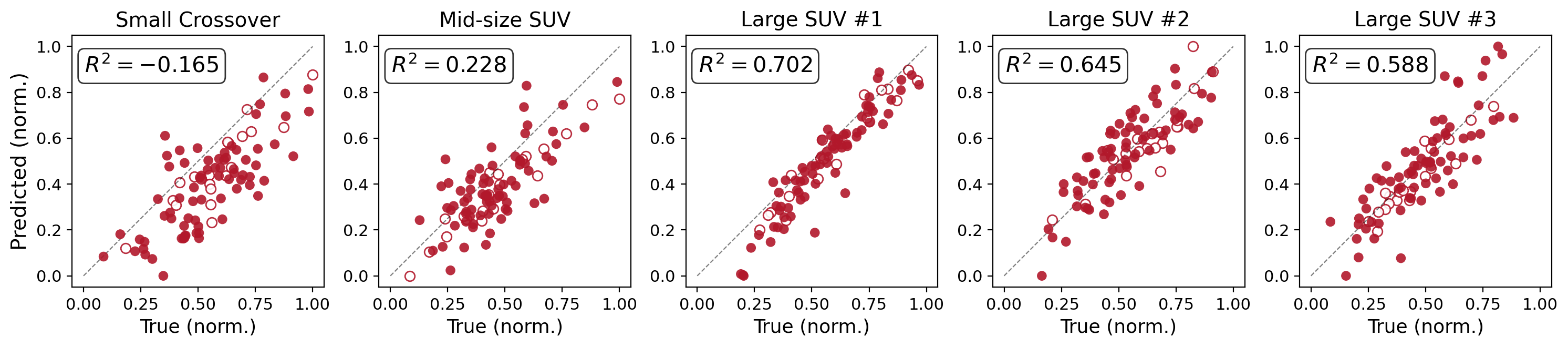}
  \caption{FFT pressure drag: predicted vs.\ ground-truth for five held-out families. Filled circles: test set; open circles: training set. Performance varies from $R^2 = 0.70$ (Large SUV \#1) to $-0.17$ (Small Crossover).}
  \label{fig:scatter_fft}
\end{figure}

Figure~\ref{fig:scatter_fft} shows the predicted vs.\ ground-truth pressure drag for each vehicle family.
For Large SUV \#1, FFT achieves a passable $R^2 = 0.70$ with visible correlation, but for Small Crossover the predictions show no meaningful structure ($R^2 = -0.17$).
Shear drag results are somewhat better (mean $R^2 = 0.71$), but the variance remains equally high, and even shear prediction for Small Crossover drops to $R^2 = 0.07$.
The unpredictable nature of these outcomes (success on some geometries, failure on others, with no way to know in advance) disqualifies FFT as a practical strategy.

\paragraph{Learning rate sensitivity analysis.}
A natural objection is that FFT's poor performance may simply reflect a suboptimal learning rate.
To rule out this possibility, we conducted a learning rate sweep over three values: $1 \times 10^{-5}$ (baseline, $\frac{1}{3}\times$ pretrain rate), $3 \times 10^{-6}$ ($\frac{1}{10}\times$), and $1 \times 10^{-6}$ ($\frac{1}{30}\times$), on two representative vehicles: Large SUV \#2 (the best-performing FFT family for pressure drag) and Small Crossover (the worst-performing).
Table~\ref{tab:fft_lr_sweep} shows that reducing the learning rate \emph{monotonically degrades} all metrics rather than improving them.

\begin{table}[ht]
  \centering
  \caption{FFT learning rate sensitivity: test-set metrics for two representative vehicles at three learning rates. All metrics degrade monotonically with lower learning rates, confirming that FFT's failure is not a hyperparameter tuning issue.}
  \label{tab:fft_lr_sweep}
  \footnotesize
  \begin{tabular}{lc cc cc cc}
    \toprule
    & & \multicolumn{2}{c}{Force $R^2$} & \multicolumn{2}{c}{Surface Pressure} & \multicolumn{2}{c}{Surface Friction} \\
    \cmidrule(lr){3-4} \cmidrule(lr){5-6} \cmidrule(lr){7-8}
    Held-out Vehicle & Learning Rate & Pressure & Shear & rel.\ $L_2$ ($\times 10^{-4}$) & NRMSE (\%) & rel.\ $L_2$ & NRMSE (\%) \\
    \midrule
    Large SUV \#2   & $1 \times 10^{-5}$ & $0.645$  & $0.950$  & $4.50$ & $1.46$ & $0.298$ & $1.83$ \\
    Large SUV \#2   & $3 \times 10^{-6}$ & $0.340$  & $0.735$  & $5.56$ & $1.80$ & $0.369$ & $2.27$ \\
    Large SUV \#2   & $1 \times 10^{-6}$ & $-1.090$ & $0.084$  & $7.17$ & $2.32$ & $0.470$ & $2.89$ \\
    \midrule
    Small Crossover & $1 \times 10^{-5}$ & $-0.165$ & $0.072$  & $5.21$ & $1.91$ & $0.405$ & $1.22$ \\
    Small Crossover & $3 \times 10^{-6}$ & $-1.295$ & $-0.229$ & $6.52$ & $2.38$ & $0.545$ & $1.65$ \\
    Small Crossover & $1 \times 10^{-6}$ & $-4.947$ & $-1.884$ & $8.38$ & $3.07$ & $0.714$ & $2.17$ \\
    \bottomrule
  \end{tabular}
\end{table}

For Large SUV \#2, reducing the learning rate from $1 \times 10^{-5}$ to $1 \times 10^{-6}$ causes pressure-drag $R^2$ to collapse from $0.645$ to $-1.09$ and surface pressure rel.\ $L_2$ to increase from $4.50$ to $7.17$; for Small Crossover, an already-failing configuration ($R^2 = -0.17$) deteriorates further to $-4.95$.
This counterintuitive result (lower learning rates producing \emph{worse} outcomes) has a clear interpretation: with only 20 training samples and 50 epochs, the optimizer receives at most 1{,}000 gradient updates.
A conservative learning rate is insufficient to traverse the high-dimensional loss landscape to reach any reasonable basin, while an aggressive rate overshoots and destabilizes.
There is no learning rate that resolves this fundamental tension for 61.47\,million unconstrained parameters trained on 20 samples.

\subsection{Lightweight Fine-Tuning}\label{sec:pft_results}

LFT freezes the geometry encoder and the first 10 surface blocks, updating only the last 2 blocks and the prediction head (6.83\,million parameters, 11.1\%).
Table~\ref{tab:pft_results} shows that this strategy fails catastrophically on pressure drag: four of five families yield $R^2 < 0$, meaning the model performs worse than simply predicting the mean.

\begin{table}[ht]
  \centering
  \caption{LFT test-set results across five leave-one-out experiments. Force $R^2$ is computed from integrated surface forces; field metrics are pointwise surface errors.}
  \label{tab:pft_results}
  \footnotesize
  \begin{tabular}{lcc cc cc}
    \toprule
    & \multicolumn{2}{c}{Force $R^2$} & \multicolumn{2}{c}{Surface Pressure} & \multicolumn{2}{c}{Surface Friction} \\
    \cmidrule(lr){2-3} \cmidrule(lr){4-5} \cmidrule(lr){6-7}
    Held-out Vehicle & Pressure & Shear & rel.\ $L_2$ ($\times 10^{-4}$) & NRMSE (\%) & rel.\ $L_2$ & NRMSE (\%) \\
    \midrule
    Small Crossover   & $-8.955$ & $-0.313$ & $10.30$ & $3.77$ & $0.785$ & $2.37$ \\
    Mid-size SUV      & $-3.641$ & $0.715$  & $6.87$  & $2.12$ & $0.454$ & $2.56$ \\
    Large SUV \#1     & $0.452$  & $0.685$  & $6.17$  & $1.64$ & $0.372$ & $1.15$ \\
    Large SUV \#2     & $-5.037$ & $0.811$  & $7.35$  & $0.75$ & $0.729$ & $1.20$ \\
    Large SUV \#3     & $-0.530$ & $0.716$  & $8.56$  & $2.78$ & $0.524$ & $3.23$ \\
    \midrule
    \textbf{Mean $\pm$ Std} & $\mathbf{-3.542 \pm 3.36}$ & $\mathbf{0.523 \pm 0.42}$ & $\mathbf{7.85 \pm 1.45}$ & $\mathbf{2.21 \pm 1.02}$ & $\mathbf{0.573 \pm 0.16}$ & $\mathbf{2.10 \pm 0.81}$ \\
    \bottomrule
  \end{tabular}
\end{table}

\begin{figure}[ht]
  \centering
  \includegraphics[width=\textwidth]{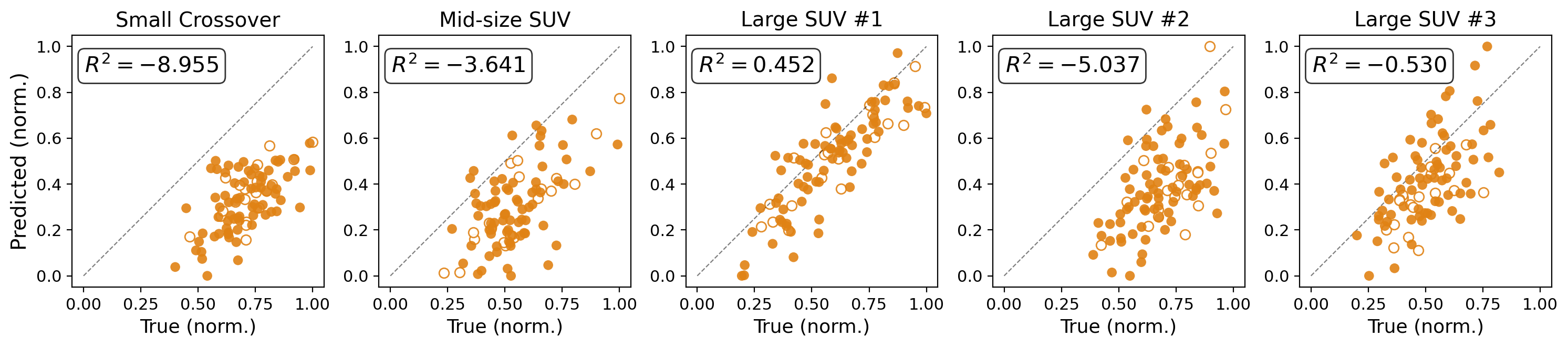}
  \caption{LFT pressure drag: predicted vs.\ ground-truth for five held-out families. Four of five families yield $R^2 < 0$. Training points (open circles) also fail, confirming fundamental inability to learn.}
  \label{fig:scatter_pft}
\end{figure}

Figure~\ref{fig:scatter_pft} confirms this visually: even the training points (open circles) deviate far from the identity line, indicating a fundamental inability to learn rather than overfitting.
The root cause is that the frozen geometry encoder produces latent representations that are uninformative for the new geometry; the downstream parameters (6.83\,million) simply cannot compensate because they never receive the correct geometric signal.

Shear drag is less catastrophic ($0.52 \pm 0.40$) because local boundary layer development, which governs shear stress, is partially captured even by a misaligned encoder, whereas pressure drag requires global geometric understanding (flow separation, wake structure) that only a properly adapted encoder can provide.
This result has a clear architectural implication: the geometry encoder \emph{must} be adapted for geometry transfer to succeed. The performance degradation in LFT compared to FFT is in good agreement with Wang~\cite{wang2025tlpinn}.
However, one key difference between Wang's study and the present study is that, in the present study, the LFT configuration is not a naive unfreezing of the final layers: the trainable components, the last two surface blocks and the prediction head, are the physically meaningful modules responsible for mapping latent representations to aerodynamic output fields.
Yet even this principled selection proves insufficient, demonstrating that the complex nonlinear physics governing pressure distribution on an unseen geometry cannot be captured by retraining a small subset of Transformer blocks alone; the geometric signal must propagate through adapted layers at every stage of the network.

\subsection{Low-Rank Adaptation}\label{sec:lora_results}

\paragraph{Rank sweep.}
To determine the optimal rank for LoRA adapters, a sweep over $r \in \{8, 16, 32, 64, 128\}$ was conducted on a single held-out family (Large SUV \#3).
As shown in Table~\ref{tab:rank_sweep}, pressure $R^2$ improves from $0.809$ at $r = 8$ to $0.852$ at $r = 16$, after which it plateaus; $r = 64$ was therefore selected for all subsequent experiments, as it achieves the best field-level accuracy (pressure rel.\ $L_2 = 2.96 \times 10^{-4}$, friction rel.\ $L_2 = 0.158$) while representing a favorable trade-off between expressiveness and parameter efficiency.
Doubling the rank to $r = 128$ doubles the trainable parameters (20.71\,M vs.\ 10.35\,M) but yields no commensurate improvement in force $R^2$ ($0.847$ vs.\ $0.852$) and only marginal field-metric gains, indicating that the model's representational bottleneck is not adapter capacity at this scale.
At this rank, each adapted linear layer adds $2 \times 64 \times 512 = 65{,}536$ parameters, and the 158 adapted layers contribute 10.13\,million total trainable parameters.

\begin{table}[ht]
  \centering
  \caption{LoRA rank sweep on the Large SUV \#3 target family. Performance plateaus at $r = 64$.}
  \label{tab:rank_sweep}
  \footnotesize
  \begin{tabular}{rc cc cc cc}
    \toprule
    & & \multicolumn{2}{c}{Force $R^2$} & \multicolumn{2}{c}{Surface Pressure} & \multicolumn{2}{c}{Surface Friction} \\
    \cmidrule(lr){3-4} \cmidrule(lr){5-6} \cmidrule(lr){7-8}
    Rank $r$ & Params (M) & Pressure & Shear & rel.\ $L_2$ ($\times 10^{-4}$) & NRMSE (\%) & rel.\ $L_2$ & NRMSE (\%) \\
    \midrule
    8   & 1.29  & $0.809$ & $0.963$ & $3.18$ & $1.04$ & $0.191$ & $1.18$ \\
    16  & 2.59  & $0.852$ & $0.971$ & $3.10$ & $1.01$ & $0.177$ & $1.09$ \\
    32  & 5.18  & $0.837$ & $0.972$ & $3.02$ & $0.99$ & $0.167$ & $1.03$ \\
    64  & 10.35 & $0.852$ & $0.972$ & $2.96$ & $0.97$ & $0.158$ & $0.98$ \\
    128 & 20.71 & $0.847$ & $0.969$ & $2.89$ & $0.94$ & $0.152$ & $0.94$ \\
    \bottomrule
  \end{tabular}
\end{table}

\paragraph{Robustness to training split.}
To assess sensitivity to the choice of training samples, a 5-fold cross-validation was conducted on the same family using the selected configuration ($r = 64$, $\alpha = 128$).
Each fold uses a different random subset of 20 training and 10 validation samples, with the remaining $\sim$73 cases for testing.
The results (Table~\ref{tab:kfold}) show low standard deviation across folds (0.041 for pressure $R^2$ and 0.005 for shear $R^2$), demonstrating that LoRA is insensitive to the specific choice of training samples.
The relatively larger force $R^2$ variance reflects the sensitivity of the scalar force metric to a small number of geometrically challenging test cases; excluding the single lowest fold (Fold~2, $R^2 = 0.748$) reduces std to 0.016, indicating that most variability is attributable to individual outlier geometries rather than systematic training instability.
Field-level metrics are even more stable (pressure rel.\ $L_2$ std $= 0.04 \times 10^{-4}$), confirming that pointwise prediction quality is robust to training split selection.
In a deployment scenario, an engineer can randomly select 20 cases for fine-tuning without needing to optimize the training split, because the rank constraint inherent to LoRA acts as an implicit regularizer that prevents overfitting to any particular subset of training points.

\begin{table}[ht]
  \centering
  \caption{5-fold cross-validation results for LoRA ($r=64$) on Large SUV \#3.}
  \label{tab:kfold}
  \footnotesize
  \begin{tabular}{c cc cc cc}
    \toprule
    & \multicolumn{2}{c}{Force $R^2$} & \multicolumn{2}{c}{Surface Pressure} & \multicolumn{2}{c}{Surface Friction} \\
    \cmidrule(lr){2-3} \cmidrule(lr){4-5} \cmidrule(lr){6-7}
    Fold & Pressure & Shear & rel.\ $L_2$ ($\times 10^{-4}$) & NRMSE (\%) & rel.\ $L_2$ & NRMSE (\%) \\
    \midrule
    0 & $0.836$ & $0.968$ & $3.03$ & $0.98$ & $0.161$ & $0.99$ \\
    1 & $0.869$ & $0.965$ & $3.03$ & $0.98$ & $0.162$ & $1.00$ \\
    2 & $0.748$ & $0.972$ & $3.00$ & $0.97$ & $0.163$ & $1.00$ \\
    3 & $0.827$ & $0.978$ & $3.13$ & $1.01$ & $0.163$ & $1.00$ \\
    4 & $0.841$ & $0.965$ & $3.06$ & $0.99$ & $0.162$ & $1.00$ \\
    \midrule
    \textbf{Mean $\pm$ Std} & $\mathbf{0.824 \pm 0.041}$ & $\mathbf{0.969 \pm 0.005}$ & $\mathbf{3.05 \pm 0.04}$ & $\mathbf{0.99 \pm 0.01}$ & $\mathbf{0.162 \pm 0.001}$ & $\mathbf{1.00 \pm 0.00}$ \\
    \bottomrule
  \end{tabular}
\end{table}

\begin{figure}[ht]
  \centering
  \includegraphics[width=\textwidth]{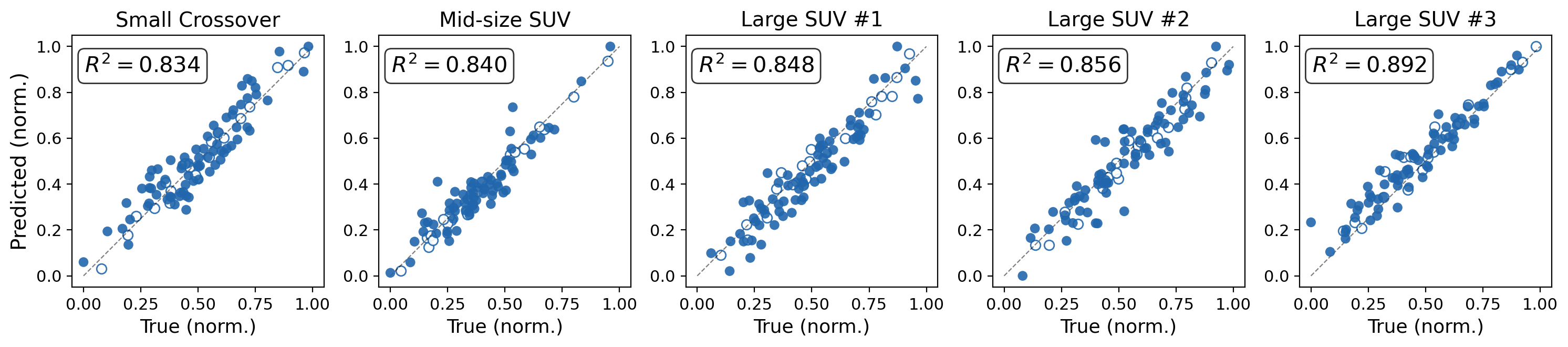}
  \caption{LoRA pressure drag: predicted vs.\ ground-truth for five held-out families. All families achieve $R^2 > 0.83$ with tight clustering around the identity line, demonstrating consistent generalization.}
  \label{fig:scatter_lora}
\end{figure}

\paragraph{Five-family results.}
Table~\ref{tab:lora_results} and Figure~\ref{fig:scatter_lora} present the per-vehicle results.
LoRA achieves pressure $R^2 = 0.851 \pm 0.020$ and shear $R^2 = 0.946 \pm 0.027$ across all five leave-one-out experiments.
The remarkably low variance indicates that the method generalizes reliably regardless of the target geometry.

Beyond integrated force accuracy, LoRA also dominates in pointwise field prediction.
The mean surface-pressure relative $L_2$ of $3.21 \times 10^{-4}$ represents a 28\,\% improvement over FFT ($4.48 \times 10^{-4}$) and a 4.4$\times$ improvement over the zero-shot baseline ($14.08 \times 10^{-4}$).
Surface friction follows the same trend: LoRA achieves rel.\ $L_2 = 0.198$ vs.\ FFT's 0.320 (38\,\% reduction).
This confirms that LoRA's advantage is not limited to the integral quantity (force) but extends to the underlying field predictions themselves.

\begin{table}[ht]
  \centering
  \caption{LoRA ($r=64$) test-set results across five leave-one-out experiments. Force $R^2$ is computed from integrated surface forces; field metrics are pointwise surface errors.}
  \label{tab:lora_results}
  \footnotesize
  \begin{tabular}{l cc cc cc}
    \toprule
    & \multicolumn{2}{c}{Force $R^2$} & \multicolumn{2}{c}{Surface Pressure} & \multicolumn{2}{c}{Surface Friction} \\
    \cmidrule(lr){2-3} \cmidrule(lr){4-5} \cmidrule(lr){6-7}
    Held-out Vehicle & Pressure & Shear & rel.\ $L_2$ ($\times 10^{-4}$) & NRMSE (\%) & rel.\ $L_2$ & NRMSE (\%) \\
    \midrule
    Small Crossover   & $0.829$ & $0.930$ & $3.08$ & $1.13$ & $0.169$ & $0.51$ \\
    Mid-size SUV      & $0.852$ & $0.969$ & $2.80$ & $0.87$ & $0.161$ & $0.90$ \\
    Large SUV \#1     & $0.837$ & $0.900$ & $3.14$ & $0.83$ & $0.158$ & $0.46$ \\
    Large SUV \#2     & $0.887$ & $0.957$ & $4.05$ & $0.36$ & $0.342$ & $0.29$ \\
    Large SUV \#3     & $0.852$ & $0.972$ & $2.96$ & $0.97$ & $0.158$ & $0.98$ \\
    \midrule
    \textbf{Mean $\pm$ Std} & $\mathbf{0.851 \pm 0.020}$ & $\mathbf{0.946 \pm 0.027}$ & $\mathbf{3.21 \pm 0.44}$ & $\mathbf{0.83 \pm 0.26}$ & $\mathbf{0.198 \pm 0.072}$ & $\mathbf{0.63 \pm 0.27}$ \\
    \bottomrule
  \end{tabular}
\end{table}

\begin{figure}[ht]
  \centering
  \begin{subfigure}[b]{0.32\textwidth}
    \includegraphics[width=\textwidth]{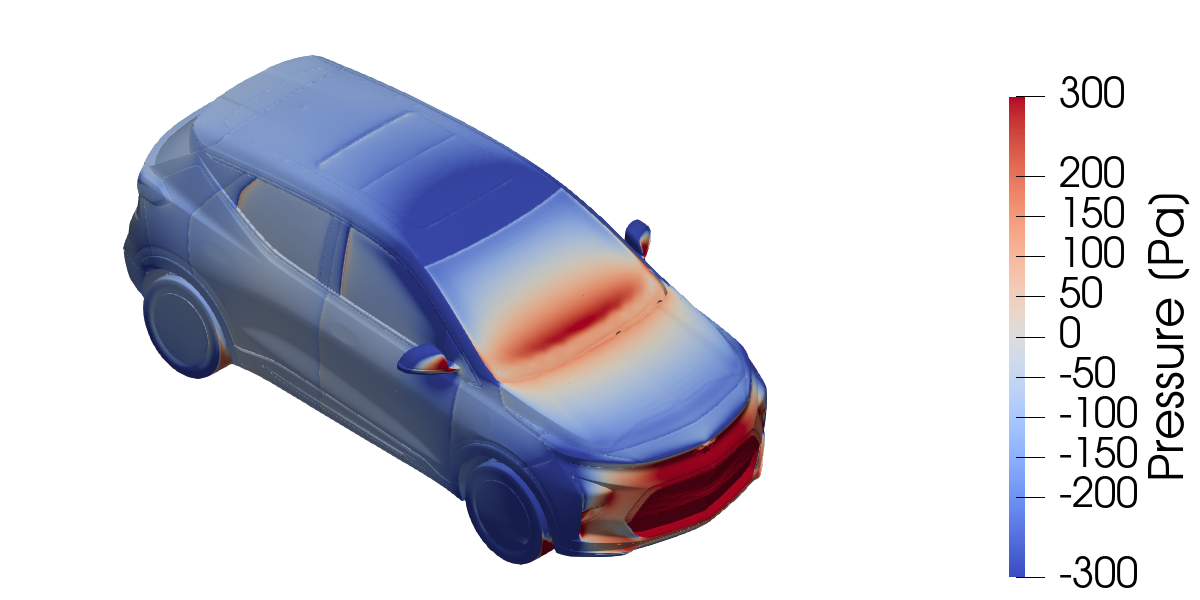}
    \caption{Small Crossover: Ground truth}
  \end{subfigure}\hfill
  \begin{subfigure}[b]{0.32\textwidth}
    \includegraphics[width=\textwidth]{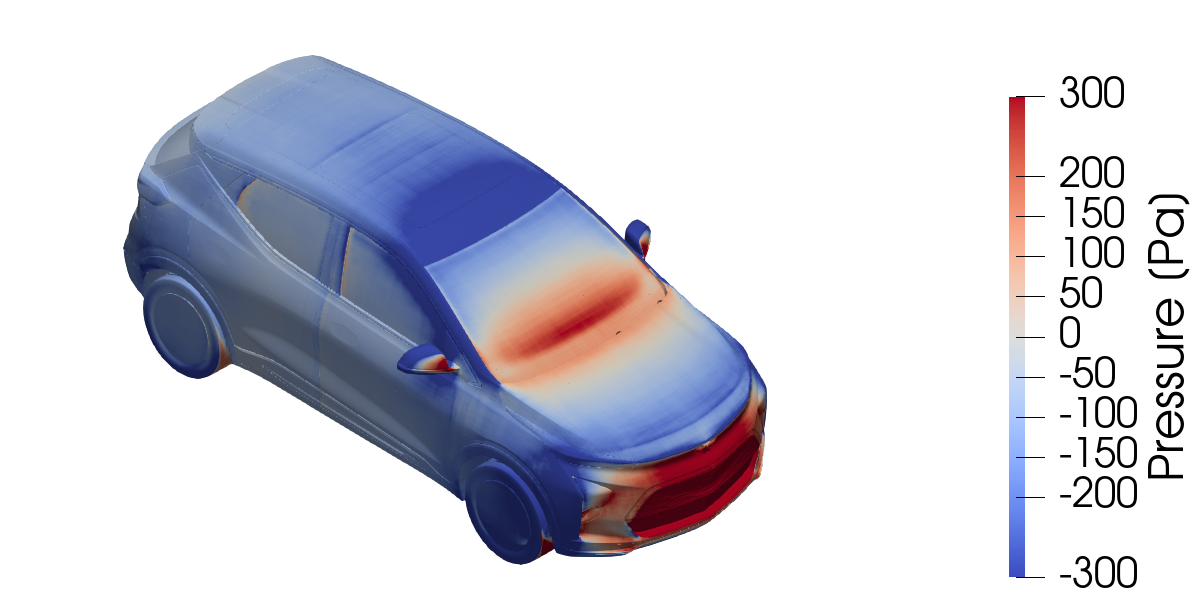}
    \caption{Small Crossover: LoRA prediction}
  \end{subfigure}\hfill
  \begin{subfigure}[b]{0.32\textwidth}
    \includegraphics[width=\textwidth]{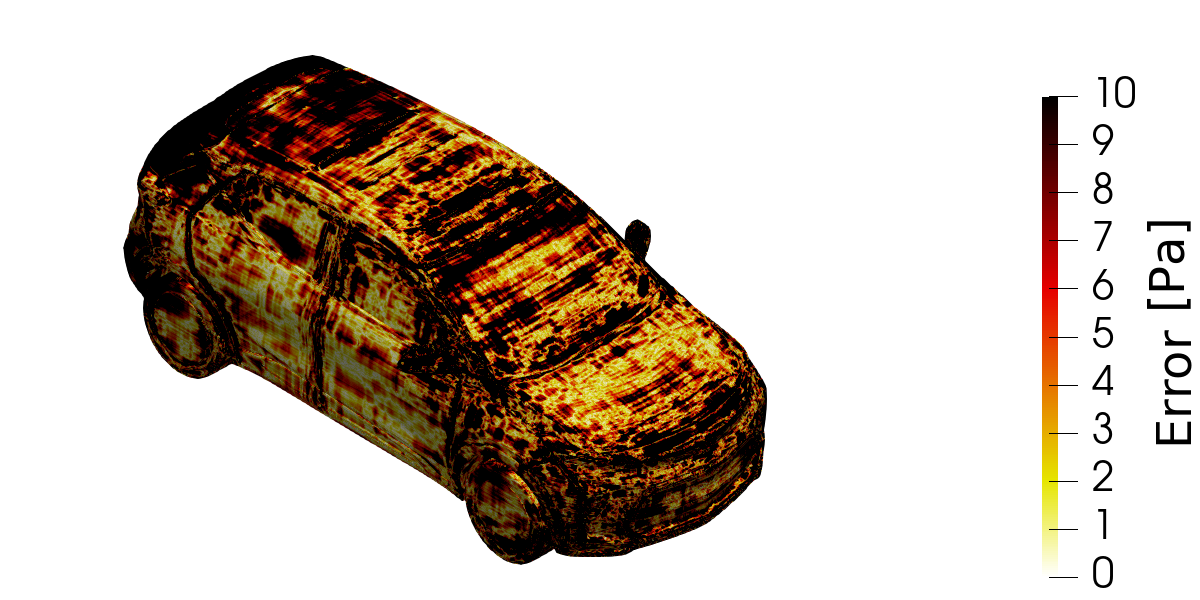}
    \caption{Small Crossover: $|$Error$|$}
  \end{subfigure}\\[4pt]
  \begin{subfigure}[b]{0.32\textwidth}
    \includegraphics[width=\textwidth]{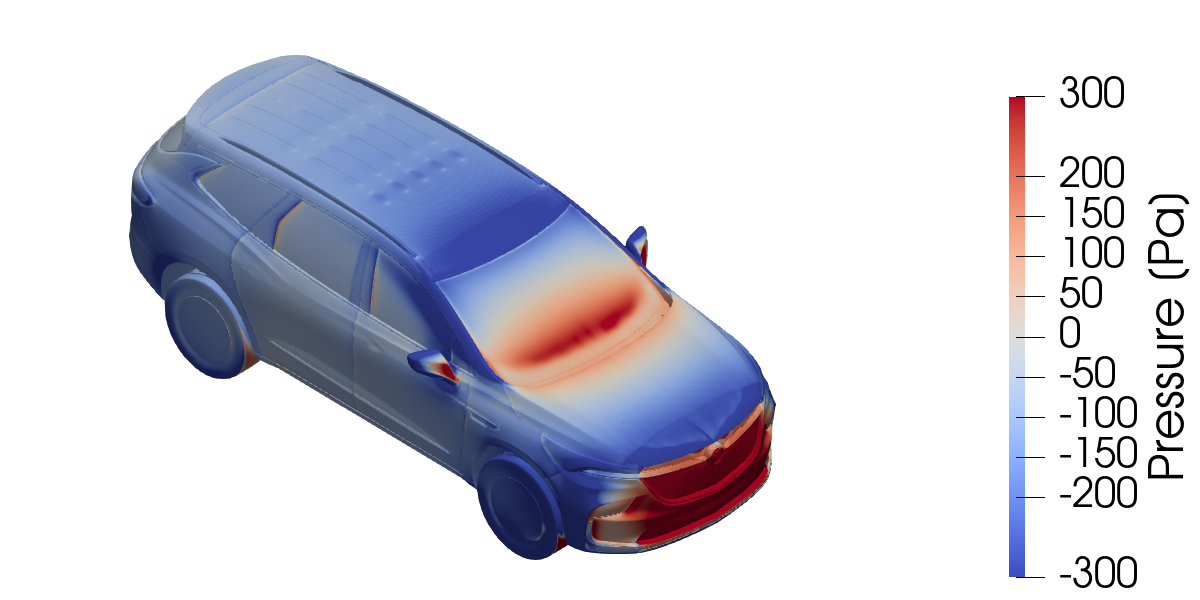}
    \caption{Large SUV \#1: Ground truth}
  \end{subfigure}\hfill
  \begin{subfigure}[b]{0.32\textwidth}
    \includegraphics[width=\textwidth]{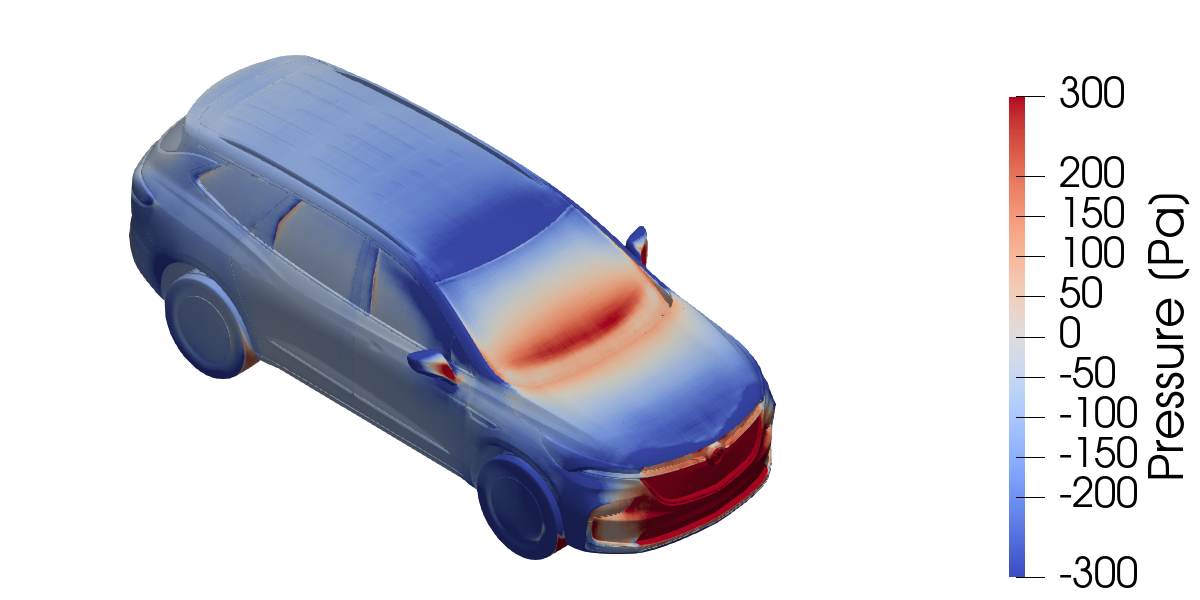}
    \caption{Large SUV \#1: LoRA prediction}
  \end{subfigure}\hfill
  \begin{subfigure}[b]{0.32\textwidth}
    \includegraphics[width=\textwidth]{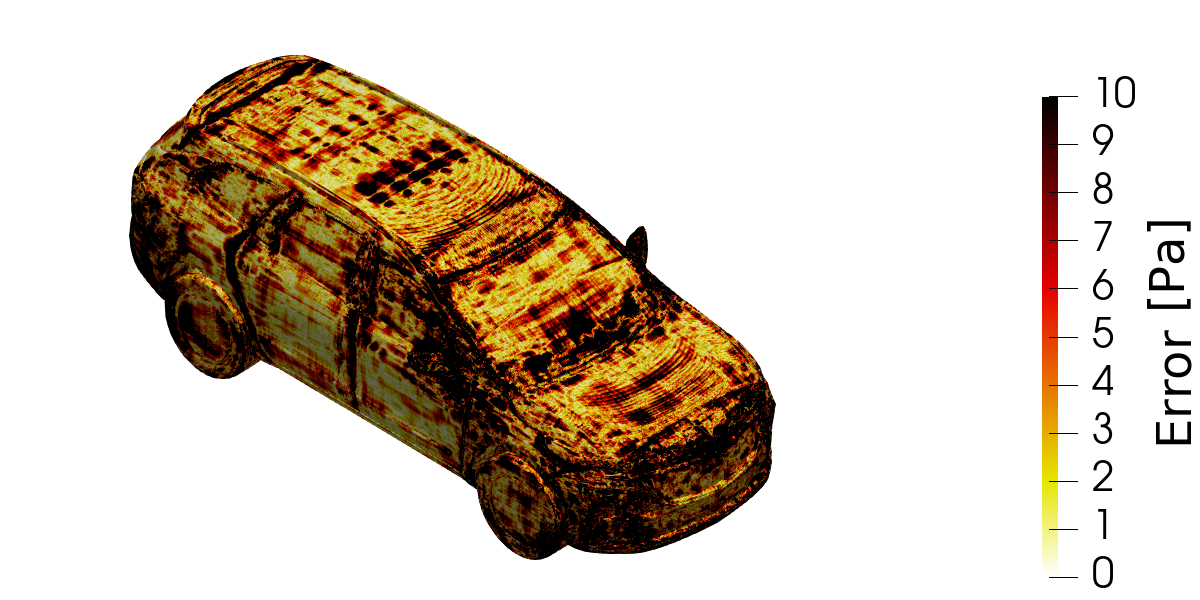}
    \caption{Large SUV \#1: $|$Error$|$}
  \end{subfigure}
  \caption{Surface pressure predictions of the LoRA-adapted model ($r=64$) on the same baseline configurations as Figure~\ref{fig:pressure_pretrain}. Unlike the pretrained model, here each vehicle is \emph{out-of-distribution}: the base checkpoint was trained without the target family and adapted with only 20 samples. Top: Small Crossover (RMSE~$= 27.8$\,Pa, drag error $= 7.5$\,N). Bottom: Large SUV \#1 (RMSE~$= 26.8$\,Pa, drag error $= 18.7$\,N). Color range: $[-300, 300]$\,Pa (pressure), $[0, 10]$\,Pa (error). Despite the harder setting, LoRA achieves comparable visual fidelity; its aggregate advantage manifests across the full DOE test distribution.}
  \label{fig:pressure_lora}
\end{figure}

Figure~\ref{fig:pressure_lora} shows the corresponding LoRA predictions on the same baseline configurations visualized in Figure~\ref{fig:pressure_pretrain}.
Critically, the comparison is asymmetric: whereas Figure~\ref{fig:pressure_pretrain} shows \emph{in-distribution} predictions from a checkpoint trained \emph{with} both vehicles, the LoRA models here start from checkpoints that have \emph{never seen} the target vehicle (leave-Small-Crossover-out and leave-Large-SUV-\#1-out, respectively) and adapt using only 20 samples.
For Small Crossover, the LoRA-adapted model achieves RMSE~$= 27.8$\,Pa (drag error $= 7.5$\,N); for Large SUV \#1, RMSE~$= 26.8$\,Pa (drag error $= 18.7$\,N).
Despite this harder out-of-distribution setting, LoRA achieves comparable visual fidelity to the in-distribution pretrained model on these well-behaved baseline geometries.
LoRA's aggregate advantage (Table~\ref{tab:lora_results}) emerges across the full test distribution including geometrically challenging Design of Experiments (DOE) variants where the pretrained model's zero-shot performance degrades substantially.

LoRA resolves both FFT's instability and LFT's failure simultaneously: by injecting low-rank adapters into \emph{all} layers including the geometry encoder, it enables the encoder to adapt to new shapes (avoiding LFT's failure), while constraining each layer's update to a rank-64 manifold (avoiding FFT's optimization instability).
The mechanism is qualitatively different from parameter reduction alone: LFT uses \emph{fewer} parameters (6.83\,million vs.\ 10.13\,million) but fails catastrophically because those parameters are concentrated in the wrong layers.
LoRA succeeds not because it has fewer parameters, but because it distributes \emph{constrained} updates across the entire network while preserving the pretrained representations as an anchor.

\subsection{Three-Way Comparison}\label{sec:three_way}

\begin{figure}[ht]
  \centering
  \includegraphics[width=0.85\textwidth]{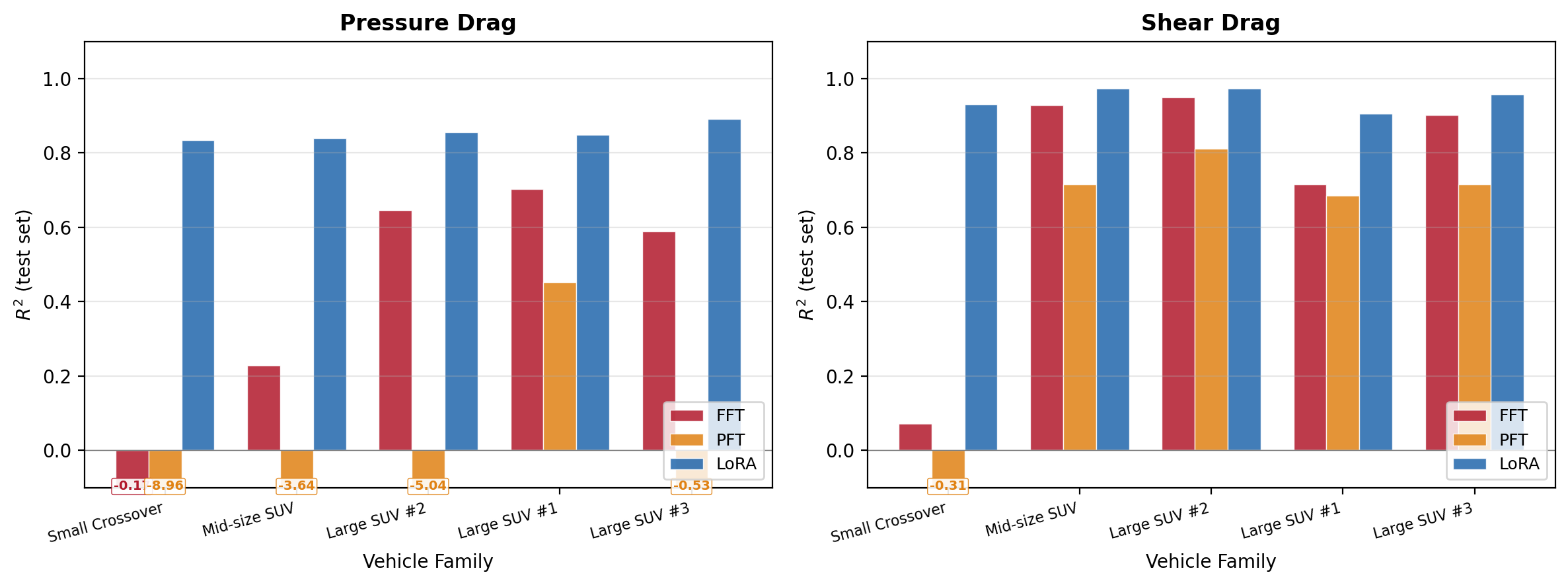}
  \caption{Per-vehicle test-set $R^2$ comparison. Left: pressure drag; Right: shear drag. LoRA (blue) dominates across all families with minimal variance.}
  \label{fig:bar_r2}
\end{figure}

Table~\ref{tab:results_combined} consolidates the mean performance, and Figure~\ref{fig:bar_r2} provides a per-vehicle comparison.

\begin{table}[ht]
  \centering
  \caption{Summary comparison: mean ($\pm$ std) test-set metrics across five leave-one-out experiments. Force $R^2$ is computed from integrated surface forces; field metrics are pointwise surface errors.}
  \label{tab:results_combined}
  \footnotesize
  \begin{tabular}{l cc cc cc}
    \toprule
    & \multicolumn{2}{c}{Force $R^2$} & \multicolumn{2}{c}{Surface Pressure} & \multicolumn{2}{c}{Surface Friction} \\
    \cmidrule(lr){2-3} \cmidrule(lr){4-5} \cmidrule(lr){6-7}
    Method & Pressure & Shear & rel.\ $L_2$ ($\times 10^{-4}$) & NRMSE (\%) & rel.\ $L_2$ & NRMSE (\%) \\
    \midrule
    FFT  & $0.400 \pm 0.33$ & $0.714 \pm 0.33$ & $4.48 \pm 0.45$ & $1.22 \pm 0.45$ & $0.319 \pm 0.070$ & $1.14 \pm 0.48$ \\
    LFT  & $-3.542 \pm 3.36$ & $0.523 \pm 0.42$ & $7.85 \pm 1.45$ & $2.21 \pm 1.02$ & $0.573 \pm 0.16$ & $2.10 \pm 0.81$ \\
    \textbf{LoRA} & $\mathbf{0.851 \pm 0.02}$ & $\mathbf{0.946 \pm 0.03}$ & $\mathbf{3.21 \pm 0.44}$ & $\mathbf{0.83 \pm 0.26}$ & $\mathbf{0.198 \pm 0.072}$ & $\mathbf{0.63 \pm 0.27}$ \\
    \bottomrule
  \end{tabular}
\end{table}

The pressure/shear asymmetry visible across all methods is physically interpretable: shear stress depends primarily on local boundary layer development, which is partially captured even by imperfect encoders, whereas pressure drag requires global geometric understanding (flow separation, wake structure) that demands a properly adapted encoder.
Since RMSE scales as $\sqrt{1 - R^2}$ for a fixed target variance, LoRA's pressure $R^2 = 0.851$ corresponds to a relative RMSE of $\sqrt{(1-0.851)/(1-0.400)} = 0.50$ compared to FFT, a 50\,\% reduction in root-mean-square prediction error.
In NLP, LoRA is primarily valued as a memory-saving tool for fine-tuning billion-parameter models; in this SciML setting, it serves as a \emph{training convergence enabler}: the rank constraint regularizes the loss landscape and prevents the catastrophic overfitting that destabilizes unconstrained fine-tuning in the extreme low-data regime.

These results contrast sharply with Wang et al.~\cite{wang2025tlpinn}, who found all three strategies performed comparably for 2D PINN geometry transfer with shallow MLPs ($\sim$30,000 parameters).
The divergence, attributable to our far greater architecture complexity and geometric dissimilarity, is analyzed in detail in Section~\ref{sec:discussion}.

\subsection{Comparison with From-Scratch Training}\label{sec:from_scratch}

\begin{figure}[ht]
  \centering
  \includegraphics[height=0.32\textheight]{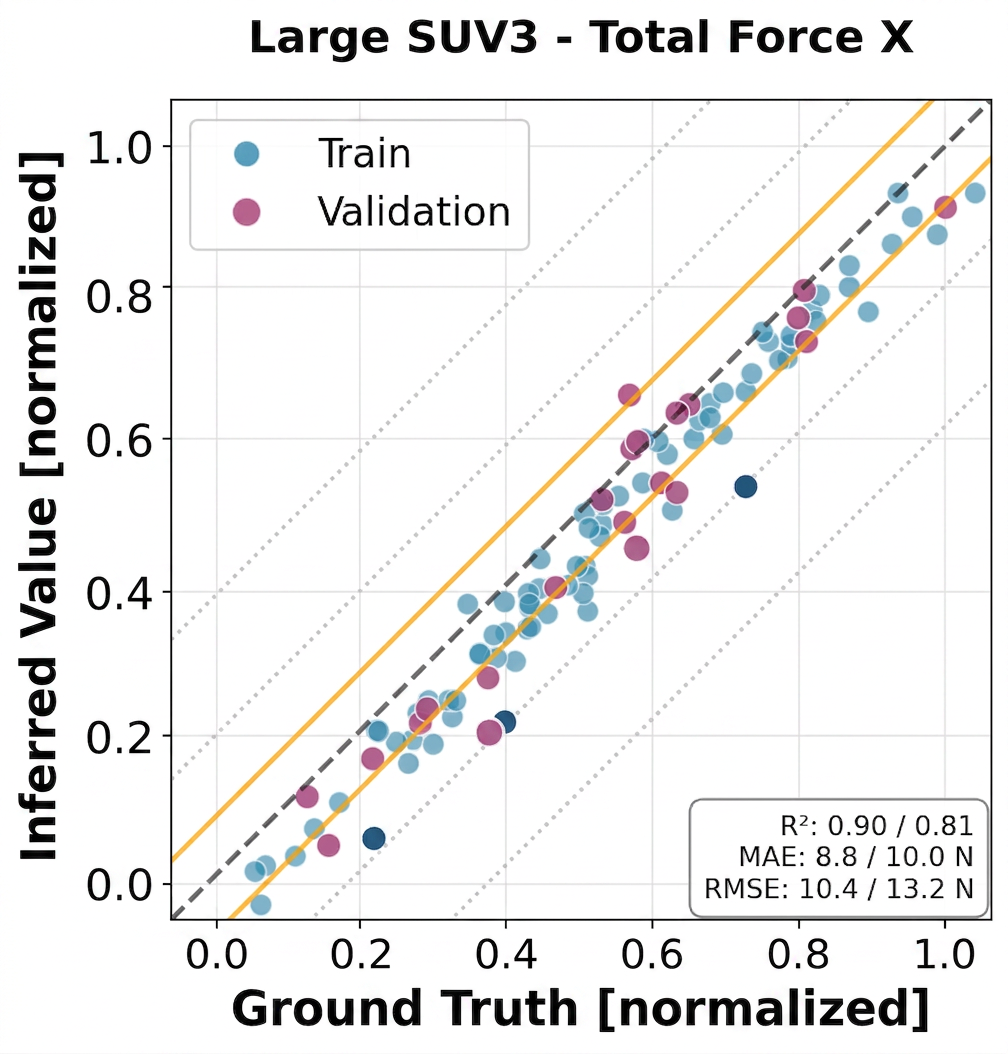}\hfill
  \includegraphics[height=0.32\textheight]{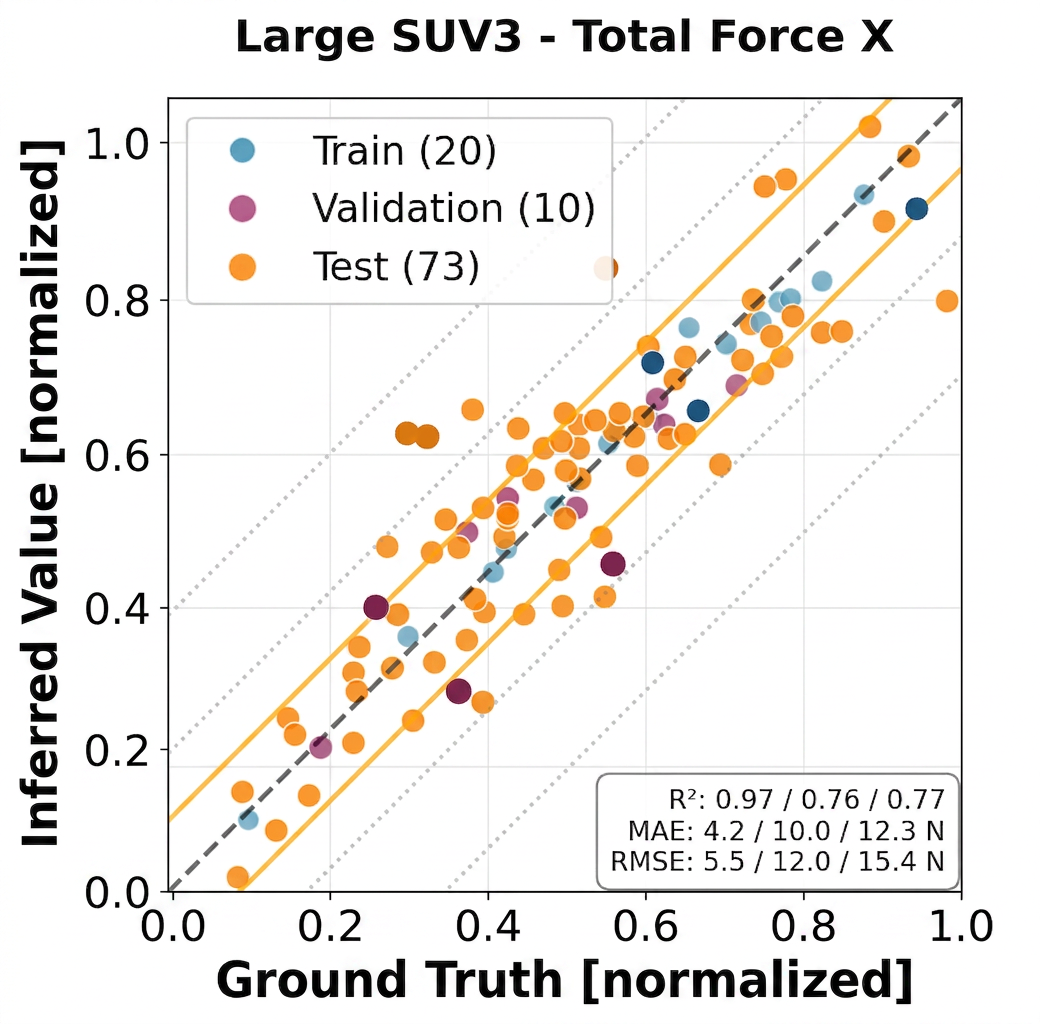}
  \caption{From-scratch training baselines for Large SUV \#3 total drag force. Left: 103 samples (train/val $R^2$ = 0.90/0.81, MAE = 10.0\,N). Right: 30 samples (test $R^2$ = 0.77, MAE = 12.3\,N).}
  \label{fig:scratch_comparison}
\end{figure}

To place the transfer learning results in broader context, we compare against models trained from scratch on all five vehicle families with the target family (Large SUV \#3) included in the training set.
Two baselines are considered: full training with 103 target-family samples (the conventional approach) and a data-limited variant with only 30 target-family samples (20 train + 10 validation).
These from-scratch models were trained identically to the pretrained model (same architecture, same hyperparameters, 500 epochs) but with the target family included from the start.
Their scatter plots for total drag force prediction are shown in Figure~\ref{fig:scratch_comparison}.

\begin{table}[ht]
  \centering
  \caption{Comparison of all training approaches for Large SUV \#3 total drag force prediction.}
  \label{tab:from_scratch}
  \small
  \begin{tabular}{lccccc}
    \toprule
    Method & Trainable & Target & Total Drag & MAE & RMSE \\
           & Params    & Data   & $R^2_\text{test}$ & (N) & (N) \\
    \midrule
    \multicolumn{6}{l}{\textit{Transfer learning (30 target samples: 20 train + 10 val)}} \\
    \textbf{LoRA} & \textbf{10.13\,M (16.5\%)} & \textbf{30} & \textbf{0.87} & \textbf{9.2} & \textbf{11.9} \\
    \midrule
    \multicolumn{6}{l}{\textit{From-scratch (target included in training)}} \\
    Full train (30) & 61.47\,M & 30 & 0.77 & 12.3 & 15.4 \\
    Full train (103) & 61.47\,M & 103 & 0.81 & 10.0 & 13.2 \\
    \bottomrule
  \end{tabular}
\end{table}

LoRA with 30 target-family samples (20 train + 10 validation) achieves total drag $R^2 = 0.87$, MAE $= 9.2$\,N, and RMSE $= 11.9$\,N, outperforming from-scratch training with 103 samples despite using more than 3$\times$ less target data and only 16.5\,\% of the trainable parameters.
The reason lies in the fundamental limitation of from-scratch multi-family training: because all vehicle families are trained jointly, the model must learn a shared representation that balances accuracy across all geometries simultaneously.
When the target family contributes few samples relative to the other families, the learned representation becomes biased toward the majority families, degrading accuracy on the under-represented target.
The performance gap between the 103-sample and 30-sample baselines ($R^2 = 0.81$ vs.\ $0.77$) directly demonstrates this effect: reducing target-family data from 103 to 30 samples (while keeping the other four families at $\sim$100 each) introduces a class imbalance that the joint optimizer cannot overcome.
To maintain accuracy on the target family in a from-scratch setting, one must ensure proportional data representation across all families, which necessitates generating a full CFD dataset for each new vehicle, a process requiring weeks of computational effort per geometry.
Training time compounds this cost: the from-scratch model requires 500 epochs over the entire dataset (511 cases, $\sim$48 GPU-hours), whereas LoRA fine-tunes only 16.5\,\% of parameters on 20 cases in approximately 5 wall-clock hours on 4$\times$A100 GPUs.
In contrast, LoRA sidesteps both limitations entirely: the pretrained encoder already captures shared geometric knowledge from the existing families, and the low-rank adapters specialize this knowledge to the target geometry without disturbing the original representations, making from-scratch retraining unnecessary.
We note that this comparison is conducted on a single target family due to the substantial computational cost of from-scratch training ($\sim$48 GPU-hours per run on 4$\times$A100-80\,GB); while the consistent LoRA superiority in the five leave-one-out transfer experiments (Section~\ref{sec:lora_results}) provides supporting evidence for generality, direct from-scratch comparisons for all families remain as future work.

\section{Discussion}\label{sec:discussion}

The results establish a clear hierarchy, LoRA $\gg$ FFT $>$ LFT for geometry transfer in the low-data regime, and this ordering carries direct implications for industrial deployment.

\paragraph{Effect of architecture and geometry complexity on transfer strategy selection.}
A central insight of this work is that the effectiveness of transfer learning strategies depends critically on the interaction between architecture complexity and geometry difficulty.
Wang et al.~\cite{wang2025tlpinn} demonstrated that for a shallow MLP ($\sim$30,000 parameters) transferring between smoothly deformed 2D geometries, all three strategies (FFT, LFT, and LoRA) perform comparably, with differences in relative error on the order of $10^{-4}$.
In that regime, the loss landscape is sufficiently well-conditioned that any reasonable optimization strategy converges to a good solution.
Our results reveal that this conclusion does not extend to more complex settings.
When the architecture scales to a multi-block Transformer with 61.47\,million parameters and cross-attention mechanisms, and the geometry transfer involves topologically distinct 3D shapes, the optimization landscape becomes fundamentally harder to navigate.
FFT, which updates all parameters freely, fails to converge stably with 20 samples; LFT, which freezes the geometry encoder entirely, fails because the encoder's representations require adaptation to encode unfamiliar geometric features.
Crucially, the learning rate sweep (Table~\ref{tab:fft_lr_sweep}) confirms that FFT's failure is not a hyperparameter issue: reducing the learning rate by up to $10\times$ monotonically degrades performance rather than improving it, demonstrating that no learning rate can resolve the fundamental ill-conditioning of 61.47\,million unconstrained parameters trained on 20 samples.
Only LoRA's rank constraint provides the necessary regularization: restricting the update to a low-dimensional subspace while still allowing all layers to adapt.
This suggests a general principle: as model architecture complexity and geometric dissimilarity increase, parameter-efficient methods transition from being a convenience (reducing memory and storage) to being a necessity (enabling convergence that would otherwise be unattainable).

The findings suggest a practical deployment paradigm: a single pretrained backbone shared across all vehicle programs, with lightweight per-family LoRA adapters trained on minimal data.
Twenty CFD simulations can typically be completed within one to two weeks of engineering effort, compared to the months required for a full training campaign; the fine-tuning itself requires approximately 5 hours on 4$\times$A100 GPUs, enabling a same-day turnaround from data to deployed model.
The adapter-based approach is modular: adding a new vehicle family requires only training a new adapter, not retraining the base model, and the low variance across training splits (K-fold std $= 0.041$) means that no careful curation of the fine-tuning dataset is required.

Several limitations should be acknowledged.
All five families are ground vehicles sharing the same physical regime (incompressible external flow at highway speeds); transfer across fundamentally different physics remains untested.
The rank $r = 64$ was selected based on a sweep for a single family, and adaptive per-layer rank selection via methods such as AdaLoRA~\cite{zhang2023adalora} or DoRA~\cite{liu2024dora} could further improve performance.
This study focuses on surface field predictions because the target engineering metric, aerodynamic drag, is computed entirely from surface quantities (pressure and wall shear stress); volume field prediction (velocity, turbulence kinetic energy) is not required for this objective.
Nevertheless, the LoRA framework is architecture-agnostic and extends naturally to combined surface--volume models; preliminary experiments with joint training are ongoing and will be reported in a follow-up study.
Finally, the current adapters are family-specific by design; investigating multi-family adapters that enable prediction on multiple unseen vehicle families simultaneously is a natural next step.

\paragraph{Normalization consistency as a hidden transfer requirement.}
A subtle but critical requirement for geometry transfer, absent in same-domain PINNs or single-geometry surrogates, is that all normalization statistics must remain anchored to the pretraining distribution.
Because the model's positional encodings, attention patterns, and learned feature scales were calibrated against a specific coordinate range and field-value distribution during pretraining, applying different normalization bounds (e.g., recomputed from the target family alone) corrupts the input space and renders the pretrained weights meaningless.
We encountered this failure mode directly during development: inadvertently recomputing coordinate normalization bounds from a different source degraded LoRA force $R^2$ from $0.86$ to $-30$ on the same checkpoint, producing predictions worse than random.
This observation generalizes to any multi-stage training pipeline in which the source and target domains differ geometrically: the normalization contract established at pretraining must be explicitly propagated through all downstream stages.
In contrast, conventional PINNs operating on fixed domains with parametric boundary conditions do not face this issue, because the spatial domain is constant and normalization is trivially consistent.

\paragraph{Zero-shot baseline and the transfer boundary.}
As established in Section~\ref{sec:pretrain_performance}, the pretrained model without any fine-tuning yields field errors approximately $4\times$ larger than the best adapted model, confirming that the encoder captures partially transferable but incomplete representations.
The contrast between zero-shot degradation, LFT's failure (encoder frozen + downstream adapted), and LoRA's success (all layers adapted at low rank) precisely delineates the boundary of what can and cannot transfer without modification: the geometry encoder must itself be updated, even at low rank, to accommodate new topologies.

\paragraph{Adapter specificity and deployment architecture.}
A practical consideration for deployment is that LoRA adapters are inherently family-specific: an adapter fine-tuned for one vehicle family degrades significantly when applied to inference on other families, because the low-rank updates specialize the model's representations toward the target geometry at the expense of the original multi-family coverage.
This finding motivates a deployment architecture based on a shared pretrained backbone with a per-family adapter registry.
At inference time, the appropriate adapter is loaded based on the input geometry, either through manual selection by the engineer or via an automated geometry-recognition module.
Since the pretrained backbone remains unchanged and each adapter contributes only $\sim$10\,million parameters ($\sim$40\,MB in storage), maintaining a registry of dozens of family-specific adapters imposes negligible overhead.

\paragraph{Toward multi-family adapters.}
The family-specificity limitation can be partially overcome by augmenting the fine-tuning dataset with a small number of samples from the pretraining families.
Preliminary experiments indicate that adding 5--10 samples per pretrained family to the fine-tuning set is sufficient to restore multi-family coverage while maintaining adaptation accuracy on the target.
This mixed-training approach points toward a Foundation Model paradigm in which a single adapter handles multiple geometries; however, fully realizing this vision requires a more diverse pretraining corpus spanning additional vehicle segments and design generations.

\paragraph{Future directions.}
Several extensions merit investigation.
First, transfer across operating conditions (e.g., vehicle speed, yaw angle) represents a natural complement to geometry transfer; a mixture-of-adapters framework such as X-LoRA~\cite{buehler2024xloramixturelowrankadapter}, in which multiple condition-specific adapters are dynamically composed at inference time, may be required to handle the combinatorial growth of adapter configurations.
Second, extending predictions from surface quantities to full volume fields (velocity, pressure, turbulence kinetic energy) will test whether the same rank settings suffice for volumetric physics.
Third, quantifying prediction uncertainty, crucial for engineering acceptance, could be addressed through Monte Carlo Dropout (MC-Dropout) ensembles~\cite{gal2016dropout} or conformal prediction~\cite{angelopoulos2023conformal} applied to the LoRA-adapted model.
Finally, connecting LoRA adapters with active learning strategies~\cite{settles2012active} could further reduce the number of CFD simulations needed, by selecting the most informative geometries for the fine-tuning set rather than sampling randomly.

\section{Conclusion}\label{sec:conclusion}

This study presented the first systematic comparison of transfer learning strategies for geometry adaptation in an industrial Transformer-based CFD surrogate.
The ultimate engineering objective is to predict the spatial distribution of surface pressure and friction force on unseen vehicle geometries, information essential for identifying drag sources and guiding design improvements, rather than merely predicting a scalar drag coefficient.
Using leave-one-family-out experiments across five topologically distinct vehicle families with only 20 fine-tuning samples, three findings emerge.

First, the geometric representations learned by the pretrained geometry encoder are transferable to unseen vehicle families.
This is evidenced by LoRA's consistent success ($R^2 = 0.851 \pm 0.02$ for pressure, $0.946 \pm 0.03$ for shear) and confirmed by LFT's failure when the encoder is frozen: the encoder contains useful but incomplete information about new shapes, requiring low-rank adaptation rather than full retraining.

Second, LoRA is not merely a parameter-efficient alternative to FFT; it is a training convergence enabler in the low-data regime.
The rank constraint restricts gradient updates to a low-dimensional subspace, regularizing the loss landscape so that stable convergence becomes possible even with 20 samples, a regime where FFT's instability ($R^2$ std $= 0.33$) and LFT's catastrophic failure ($R^2 = -3.54$ mean) demonstrate that unconstrained or layer-restricted updates are not viable.

Third, LoRA with 30 target-family samples (20 train + 10 validation) outperforms from-scratch training with 103 target-family samples ($R^2 = 0.87$ vs.\ $0.81$ for total drag), representing more than a 3$\times$ reduction in target-family data requirements.
Combined with the low sensitivity to training split selection (K-fold std $= 0.041$), this establishes a practical pathway: 20 randomly selected simulations suffice for reliable geometry adaptation.

A prerequisite underpinning all three findings is strict normalization consistency: the coordinate and field statistics established during pretraining must be propagated unchanged through fine-tuning and inference; violating this contract renders the pretrained weights unusable regardless of the adaptation strategy employed.

These results point toward a deployment paradigm of shared pretrained backbones with per-family LoRA adapters, enabling rapid adaptation to new vehicle families with minimal engineering data and compute, and removing a key barrier to the industrial adoption of SciML surrogates.
The adapter-based architecture is inherently modular: adding a new vehicle family requires only training a new adapter ($\sim$5 GPU-hours, 20 CFD cases), not retraining the base model, and the approach extends naturally to transfer across operating conditions and to full-field predictions.

\bibliographystyle{unsrtnat}
\bibliography{references}

\end{document}